\documentclass[12pt]{article}
\begin{document}
\title{Constitutive equations for a polymer fluid
based on the concept of non-affine networks}

\author{A.D. Drozdov\footnote{Corresponding author.
Fax: (304) 293 4139; E-mail: Aleksey.Drozdov@mail.wvu.edu}
\hspace*{1 mm} and R.K. Gupta\\
Department of Chemical Engineering\\
West Virginia University\\
P.O. Box 6102\\
Morgantown, WV 26506, USA}
\date{}
\maketitle

\begin{abstract}
Constitutive equations are developed for a polymer
fluid, which is treated as a permanent network of
strands bridged by junctions.
The junctions are assumed to slide with respect
to their reference positions under loading.
Governing equations are derived by using the laws
of thermodynamics under the assumption that the
vorticity tensor for the flow of junctions is
proportional to that for macro-deformation.
Explicit expressions are developed for the steady
elongational viscosity, as well as for the steady
shear viscosity and normal stress functions.
To verify the constitutive relations, three sets of
experimental data are approximated on polystyrene
solutions with various molecular weights.
It is demonstrated that the model can correctly
describe stress overshoot for the shear stress
and first normal stress difference in start-up
tests with various strain rates.
Adjustable parameters in the governing equations
change consistently with the strain rate,
molecular weight and concentration of entanglements.
To validate the constitutive equations,
observations on low-density polyethylene melt
in uniaxial extensional flow are compared with
the results of numerical analysis when
the material constants are found by matching
experimental data in shear tests.
\end{abstract}


\section{Introduction}

This paper is concerned with modelling the rate-dependent
response of polymer fluids at finite strains.
In particular, the present study focuses on
constitutive equations that can adequately describe
stress overshoot under shear.
The latter phenomenon has attracted substantial attention
in the past three decades, because it provides a severe
test for the analysis of rheological models.

Experimental studies demonstrate that under shear flow
with a constant strain rate at isothermal conditions,
the shear stress and the first normal stress difference
grow monotonically with time to their ultimate
values when the shear rate is relatively small.
With an increase in the strain rate, the dependence of
the shear stress on time becomes non-monotonic:
it reaches a maximum at some instant (overshoot)
and decreases to its limiting value afterwards
(strain softening).
At sufficiently high shear rates, the dependencies
of both the shear stress and the first normal stress
difference on time becomes non-monotonic,
but the time when the first normal stress difference
reaches its maximum noticeably exceeds that for the shear
stress.
The instants when the shear stress and the first normal
stress difference become maximal are strongly
affected by strain rate: the higher the shear rate is,
the earlier the overshoots occur and the larger are
the maximal stresses.

Stress overshoot has been observed in conventional
shear tests on polymer melts
\cite{Mei72,Wag76,Phi77,SBB82,RW00,SR01},
immiscible molten blends of polymers \cite{LGC98},
polymer melts filled with micro- \cite{BP85}
and nano-particles \cite{SAS01,Dro04},
polymer solutions \cite{OLM98,OII00,OIU00a},
solutions of associative copolymers \cite{KCS95},
surfactant solutions \cite{FR97},
solutions of wormlike micelles \cite{Ber97,LDH98},
liquid crystalline polymers \cite{HUB01}
and discotic mesophases \cite{GR03},
bicontinuous microemulsions \cite{KBW02},
colloidal suspensions \cite{YCM99,MBA02},
and suspensions of rod-like macromolecules \cite{SC87},
to mention a few.
This list demonstrates that stress overshoot is a rather
common phenomenon observed in complex fluids.
Comparison of experimental data for polymer solutions
and melts with results of numerical simulation reveals
that several constitutive equations can
qualitatively describe stress overshoot under shear.
We would mention among them the single integral
model \cite{Wag76},
the Giesekus model \cite{Gie82},
the concept of temporary networks \cite{TE92},
the finitely extensible nonlinear elastic (FENE) network
model \cite{GW97},
the reptation concept that incorporates segmental stretching
\cite{OLM98,HSJ99},
and the pom--pom model \cite{BMH00,PL01}.

However, even the most sophisticated constitutive equations
that include a number of adjustable parameters
can reproduce observations
for the shear stress and the first normal stress difference
in start-up shear tests only qualitatively,
and, when their parameters are fitted to describe the stress
overshoot, they fail to adequately predict the steady shear
viscosity (see Figure 12 in \cite{PL01}).
This observation may be explained by the fact that
physical reasons for the origin of stress overshoot
remain rather obscure, and not all of them are taken
into account in conventional models.
The following possible mechanisms for the onset of
stress overshoot are mentioned:
(i) changes in the concentration of entanglements
and entanglement spacing \cite{KCS95},
(ii) changes in the orientation of polymer coils
\cite{Sch93}
and nonuniform segmental stretching \cite{Sch93},
(iii) formation of an anisotropic mechanically-induced
mesoscopic structure \cite{SAS01}
that becomes unstable at relatively high strain rates
\cite{Ber97},
and (iv) reorientation of meso-domains toward a
steady-state aligned flow \cite{HUB01,GR03}.

Although it is conventionally accepted that ordering
of strands and their reorientation driven by shear
flow provide the most important mechanism for stress
overshoot and subsequent strain softening,
to the best of our knowledge,
no attempts have been made to incorporate this phenomenon
into constitutive equations explicitly and to assess
the effect of molecular structure of a polymer
on the kinetics of reorientation.
An important point here is that at finite strains,
the evolution of an infinitesimal volume that can
rotate with respect to macro-deformations
is determined by two tensors: a symmetric
rate-of-strain tensor and a skew-symmetric vorticity
tensor (the so-called plastic spin),
while the latter is conventionally disregarded in
constitutive models for polymer fluids
(despite the observation that ``it is the vorticity in
shearing that leads to its extreme strain softening"
\cite{ML98}).

The objective of this study is two-fold:
(i) to develop constitutive equations for the
response of polymer fluids that account for
the evolution of the plastic spin,
on the one hand, and that involve a relatively small
number of material constants, on the other,
and (ii) to demonstrate that these relations describe
stress overshoot under shear of polymer solutions
and melts quantitatively,
while their adjustable parameters change consistently
with strain rate and molecular structure of polymer fluids.

A polymer fluid is treated as a network
of chains bridged by junctions (entanglements and
physical cross-links whose life-time exceeds the
characteristic time of an experiment).
The network is assumed to move non-affinely: the
junctions between strands can slide with respect
to their reference positions in the bulk material
under deformation.
Following \cite{JS77}, we suppose that the vorticity
tensor for viscoplastic flow of junctions is proportional
to that for macro-deformation, whereas the rate-of-strain
tensor for sliding of junctions is determined by the laws
of thermodynamics.
To simplify the analysis, we disregard rearrangement
of strands in the network (which implies that
we confine ourselves to relatively slow motions
whose characteristic time exceeds the characteristic
time for relaxation of stresses).
This allows constitutive equations to be derived
which describe any set of observations for the shear
stress and the first normal stress difference in
start-up shear tests by only four constants that
have a transparent physical meaning.
These parameters are found by matching experimental
data for monodisperse polystyrene (PS) solutions with
various mass-average molecular weights and molecular
weights between entanglements and for a melt of
low-density polyethylene (LDPE).

The exposition is organized as follows.
Basic kinematic relations for a non-affine network
are derived in Section 2.
Constitutive equations for a polymer fluid are
developed in Section 3, where it is demonstrated
that these relations generalize some well-known
rheological models
(Johnson--Segalman model \cite{JS77},
Phan-Thien--Tanner model \cite{PTT77},
Giesekus model \cite{Gie82},
and Leonov model \cite{SL95}).
Phenomenological relations for adjustable functions
in the governing equations are suggested in Section 4.
The governing relations are simplified for
uniaxial extension in Section 5 and simple shear
in Section 6, where explicit formulas are developed
for the steady elongational and shear viscosities.
In Section 7, material constants are found by
fitting observations in start-up shear tests.
A discussion of the effects of shear rate
and molecular weight distribution on these quantities
is provided in Section 8.
The model is validated in Section 9 by comparison of
results of numerical simulation for LDPE melt in
extensional and shear flows.
Some concluding remarks are formulated in Section 10.

\section{Kinematic relations}

With reference to the concept of non-affine networks,
a polymer fluid is modelled as a network of strands
bridged by junctions that can slide with respect
to their reference positions under deformation.
Denote by ${\bf F}(t)$ the deformation gradient for
transition from the initial state to the actual state
at time $t\geq 0$, and by ${\bf F}_{\rm s}(t)$ the
deformation gradient for transition from the initial
state to the reference state associated with sliding
of junctions.
It is presumed that the transformations described by
the tensors ${\bf F}(t)$ and ${\bf F}_{\rm s}(t)$ are
volume-preserving, which means that the third principal
invariants of these tensors equal unity.

The elastic deformation gradient ${\bf F}_{\rm e}(t)$
[the deformation gradient for transition from the
current reference state for junctions (the so-called
intermediate state) to the actual state of the network]
is determined by the conventional formula for
the multiplicative decomposition of the deformation
gradient
\begin{equation}
{\bf F}_{\rm e}(t)={\bf F}(t)\cdot {\bf F}_{\rm s}^{-1}(t),
\end{equation}
where the dot stands for inner product.
Differentiation of Eq. (1) with respect to time $t$
implies that
\begin{equation}
\frac{d{\bf F}_{\rm e}}{dt}(t)
={\bf L}(t)\cdot {\bf F}_{\rm e}(t)
-{\bf F}(t)\cdot {\bf F}_{\rm s}^{-1}(t)
\cdot \frac{d{\bf F}_{\rm s}}{dt}(t)\cdot
{\bf F}_{\rm s}^{-1}(t),
\end{equation}
where
\begin{equation}
{\bf L}(t)=\frac{d{\bf F}}{dt}(t)\cdot {\bf F}^{-1}(t)
\end{equation}
is the velocity gradient.
Using Eq. (1) and introducing the elastic velocity gradient
${\bf L}_{\rm e}(t)$ and the velocity gradient for sliding
of junctions ${\bf l}_{\rm s}(t)$
by the formulas analogous to Eq. (3),
\begin{equation}
{\bf L}_{\rm e}(t)=\frac{d{\bf F}_{\rm e}}{dt}(t)
\cdot {\bf F}_{\rm e}^{-1}(t),
\qquad
{\bf l}_{\rm s}(t)=\frac{d{\bf F}_{\rm s}}{dt}(t)
\cdot {\bf F}_{\rm s}^{-1}(t),
\end{equation}
we present Eq. (2) in the form
\begin{equation}
{\bf L}_{\rm e}(t)={\bf L}(t)-{\bf F}_{\rm e}(t)
\cdot {\bf l}_{\rm s}(t)\cdot {\bf F}_{\rm e}^{-1}(t).
\end{equation}
Setting
\begin{equation}
{\bf L}_{\rm s}(t)={\bf F}_{\rm e}(t)
\cdot {\bf l}_{\rm s}(t)\cdot {\bf F}_{\rm e}^{-1}(t)
\end{equation}
in Eq. (5), we arrive at the additive decomposition
of the velocity gradients
\begin{equation}
{\bf L}_{\rm e}(t)={\bf L}(t)-{\bf L}_{\rm s}(t).
\end{equation}
The left and right Cauchy--Green tensors for elastic
deformation are given by
\begin{equation}
{\bf B}_{\rm e}(t) = {\bf F}_{\rm e}(t)\cdot
{\bf F}_{\rm e}^{\top}(t),
\qquad
{\bf C}_{\rm e}(t) = {\bf F}_{\rm e}^{\top}(t)
\cdot {\bf F}_{\rm e}(t),
\end{equation}
where $\top$ denotes transpose.
We differentiate the first equality in Eqs. (8) with
respect to time, use Eqs. (4) and (7), and find that
\begin{eqnarray}
\frac{d{\bf B}_{\rm e}}{dt}(t)
&=& \Bigl [{\bf L}(t)\cdot {\bf B}_{\rm e}(t)
+{\bf B}_{\rm e}(t)\cdot {\bf L}^{\top}(t)\Bigr ]
-\Bigl [ {\bf L}_{\rm s}(t)\cdot {\bf B}_{\rm e}(t)
+{\bf B}_{\rm e}(t)\cdot {\bf L}_{\rm s}^{\top}(t)\Bigr ].
\end{eqnarray}
Differentiation of the other equality in Eqs. (8)
results in
\begin{equation}
\frac{d{\bf C}_{\rm e}}{dt}(t)
=2 {\bf F}_{\rm e}^{\top}(t)\cdot {\bf D}_{\rm e}(t)
\cdot {\bf F}_{\rm e}(t),
\end{equation}
where
\[
{\bf D}_{\rm e}(t)=\frac{1}{2}\Bigl [ {\bf L}_{\rm e}(t)
+{\bf L}_{\rm e}^{\top}(t) \Bigr ]
\]
is the rate-of-strain tensor for elastic deformation.
It follows from Eqs. (8) and (10) that
\begin{equation}
\frac{d{\bf C}_{\rm e}^{-1}}{dt}(t)
=-2 {\bf F}_{\rm e}^{-1}(t)\cdot {\bf D}_{\rm e}(t)
\cdot {\bf F}_{\rm e}^{-\top}(t).
\end{equation}
For volume-preserving deformations,
the first and second principal invariants
of the right Cauchy--Green tensor ${\bf C}_{\rm e}(t)$
read
\begin{equation}
J_{\rm e1}(t)={\cal I}_{1}\Bigl ({\bf C}_{\rm e}(t)\Bigr )
={\bf C}_{\rm e}(t):{\bf I},
\qquad
J_{\rm e2}(t)={\cal I}_{1}\Bigl ({\bf C}_{\rm e}^{-1}(t)\Bigr )
={\bf C}_{\rm e}^{-1}(t):{\bf I}.
\end{equation}
where ${\cal I}_{1}$ denotes the first invariant
of a tensor,
${\bf I}$ is the unit tensor,
and the colon stands for convolution.
Differentiating the first equality in Eqs. (12)
with respect to time and using Eqs. (8) and (10),
we obtain
\begin{equation}
\frac{dJ_{\rm e1}}{dt}(t)
=2 {\bf B}_{\rm e}(t):{\bf D}_{\rm e}(t).
\end{equation}
Differentiation of the other equality in Eqs. (12)
together with Eqs. (8) and (11) results in
\begin{equation}
\frac{dJ_{\rm e2}}{dt}(t)
=-2 {\bf B}_{\rm e}^{-1}(t):{\bf D}_{\rm e}(t).
\end{equation}
Equations (13) and (14) imply that the derivative of
an arbitrary smooth function $\Phi(J_{\rm e1},J_{\rm e2})$
of the first two principal invariants of the right
Cauchy--Green tensor for elastic deformation
${\bf C}_{\rm e}(t)$ with respect to time $t$ is given by
\begin{equation}
\frac{d\Phi}{dt}\Bigl (J_{\rm e1}(t),J_{\rm e2}(t)\Bigr ) =
2 \Bigl [ \Phi_{1}(t){\bf B}_{\rm e}(t)
-\Phi_{2}(t) {\bf B}_{\rm e}^{-1}(t) \Bigr ]:{\bf D}_{\rm e}(t),
\end{equation}
where
\[
\Phi_{m}(t)=\frac{\partial \Phi}{\partial J_{{\rm e}m}}
\Bigl (J_{\rm e1}(t),J_{\rm e2}(t)\Bigr )
\qquad
(m=1,2).
\]
It follows from Eq. (7) that
\begin{equation}
{\bf D}_{\rm e}(t)={\bf D}(t)-{\bf D}_{\rm s}(t),
\end{equation}
where
\[
{\bf D}(t)=\frac{1}{2}\Bigl [ {\bf L}(t)
+{\bf L}^{\top}(t) \Bigr ],
\qquad
{\bf D}_{\rm s}(t)=\frac{1}{2}\Bigl [ {\bf L}_{\rm s}(t)
+{\bf L}_{\rm s}^{\top}(t) \Bigr ]
\]
are the rate-of-strain tensors for macro-deformation
and sliding of junctions, respectively.
Combining Eqs. (15) and (16), we find that
\begin{eqnarray}
\frac{d\Phi}{dt}\Bigl (J_{\rm e1}(t),J_{\rm e2}(t)\Bigr ) &=&
2 \biggl \{ \Bigl [ \Phi_{1}(t){\bf B}_{\rm e}(t)
-\Phi_{2}(t) {\bf B}_{\rm e}^{-1}(t) \Bigr ]:{\bf D}(t)
\nonumber\\
&& -\Bigl [ \Phi_{1}(t){\bf B}_{\rm e}(t)
-\Phi_{2}(t) {\bf B}_{\rm e}^{-1}(t) \Bigr ]:{\bf D}_{\rm s}(t)
\biggr \}.
\end{eqnarray}
It is convenient to present the velocity gradients
${\bf L}(t)$ and ${\bf L}_{\rm s}(t)$ as the sums
of the rate-of-strain tensors and vorticity tensors,
\begin{equation}
{\bf L}(t)={\bf D}(t)+{\bf \Omega}(t),
\qquad
{\bf L}_{\rm s}(t)={\bf D}_{\rm s}(t)+{\bf \Omega}_{\rm s}(t),
\end{equation}
where
\[
{\bf \Omega}(t)=\frac{1}{2}\Bigl [ {\bf L}(t)
-{\bf L}^{\top}(t) \Bigr ],
\qquad
{\bf \Omega}_{\rm s}(t)=\frac{1}{2}\Bigl [
{\bf L}_{\rm s}(t) -{\bf L}_{\rm s}^{\top}(t) \Bigr ].
\]
Without loss of generality, we present the rate-of-strain
tensor ${\bf D}_{\rm s}(t)$ in the form
\begin{equation}
{\bf D}_{\rm s}(t)=\beta(t) {\bf D}(t)+{\bf M}(t),
\end{equation}
where $\beta(t)$ is a non-negative scalar function
and
\begin{equation}
{\bf M}(t)={\bf D}_{\rm s}(t)-\beta(t) {\bf D}(t).
\end{equation}
Substitution of expression (19) into Eq. (17) implies
that
\begin{eqnarray}
\frac{d\Phi}{dt}\Bigl (J_{\rm e1}(t),J_{\rm e2}(t)\Bigr ) &=&
2 \biggl \{ \Bigl (1-\beta(t)\Bigr )
\Bigl [ \Phi_{1}(t){\bf B}_{\rm e}(t)
-\Phi_{2}(t) {\bf B}_{\rm e}^{-1}(t) \Bigr ]^{\prime}:{\bf D}(t)
\nonumber\\
&& -\Bigl [ \Phi_{1}(t){\bf B}_{\rm e}(t)
-\Phi_{2}(t) {\bf B}_{\rm e}^{-1}(t) \Bigr ]^{\prime}:{\bf M}(t)
\biggr \},
\end{eqnarray}
where the prime stands for the deviatoric component of a
tensor and we bear in mind that the tensors ${\bf D}(t)$
and ${\bf M}(t)$ are traceless.
Following \cite{JS77}, we assume the vorticity tensor
for sliding of junctions to be proportional to
the vorticity tensor for macro-deformation
\begin{equation}
{\bf \Omega}_{\rm s}(t)=\gamma(t) {\bf \Omega}(t),
\end{equation}
where $\gamma(t)$ is a non-negative scalar function.
Equations (18), (19) and (22) result in
\begin{equation}
{\bf L}_{\rm s}(t)={\bf M}(t)+\beta(t){\bf D}(t)
+\gamma(t){\bf \Omega}(t).
\end{equation}
Combining Eqs. (9) and (22), we arrive at the differential
equation for the left Cauchy--Green tensor for elastic
deformation
\begin{eqnarray*}
\frac{d{\bf B}_{\rm e}}{dt}(t)
&=& \Bigl [{\bf L}(t)\cdot {\bf B}_{\rm e}(t)
+{\bf B}_{\rm e}(t)\cdot {\bf L}^{\top}(t)\Bigr ]
-\beta(t) \Bigl [ {\bf D}(t)\cdot {\bf B}_{\rm e}(t)
+{\bf B}_{\rm e}(t)\cdot {\bf D}(t)\Bigr ]
\nonumber\\
&&-\Bigl [ {\bf M}(t)\cdot {\bf B}_{\rm e}(t)
+{\bf B}_{\rm e}(t)\cdot {\bf M}(t)\Bigr ]
-\gamma(t) \Bigl [ {\bf \Omega}(t)\cdot {\bf B}_{\rm e}(t)
-{\bf B}_{\rm e}(t)\cdot {\bf \Omega}(t)\Bigr ],
\end{eqnarray*}
where we take into account that ${\bf M}(t)$ is a
symmetric tensor and ${\bf \Omega}(t)$ is a
skew-symmetric tensor.
Excluding the vorticity tensor ${\bf \Omega}(t)$
by means of Eq. (18), we obtain
\begin{eqnarray}
\frac{d{\bf B}_{\rm e}}{dt}(t)
&=& \Bigl (1-\gamma(t) \Bigr )
\Bigl [{\bf L}(t)\cdot {\bf B}_{\rm e}(t)
+{\bf B}_{\rm e}(t)\cdot {\bf L}^{\top}(t)\Bigr ]
\nonumber\\
&& +\Bigl (\gamma(t)-\beta(t)\Bigr )
\Bigl [ {\bf D}(t)\cdot {\bf B}_{\rm e}(t)
+{\bf B}_{\rm e}(t)\cdot {\bf D}(t)\Bigr ]
\nonumber\\
&&-\Bigl [ {\bf M}(t)\cdot {\bf B}_{\rm e}(t)
+{\bf B}_{\rm e}(t)\cdot {\bf M}(t)\Bigr ].
\end{eqnarray}
Our aim now is to apply Eqs. (21) and (24) to derive
constitutive equations for a polymer fluid by using
the laws of thermodynamics.

\section{Constitutive equations}

Denote by $N$ the average number of strands per unit
volume of a network and by $w$ the average strain
energy per strand.
For an isotropic incompressible network, the quantity
$w$ is treated as a function of the first two principal
invariants of the right Cauchy--Green tensor,
$w=w \Bigl (J_{\rm e1}(t), J_{\rm e2}(t)\Bigr )$.
Neglecting the energy of interaction between strands
(this energy is accounted for by means of the
incompressibility condition \cite{TE92}), we define
the strain energy per unit volume as the sum of
the strain energies of strands,
\begin{equation}
W=Nw.
\end{equation}
For isothermal deformation of an incompressible medium
at a reference temperature $\Theta_{0}$,
the Clausius--Duhem inequality reads
\[
Q(t)=-\frac{dW}{dt}(t)
+{\bf \Sigma}^{\prime}(t):{\bf D}(t)\geq 0,
\]
where $Q$ is internal dissipation per unit volume,
and ${\bf \Sigma}$ is the Cauchy stress tensor.
Substituting expression (25) into this formula
and using Eq. (21), we find that
\begin{eqnarray}
Q(t) &=& -2 N \biggl \{ \Bigl (1-\beta(t)\Bigr )
\Bigl [ w_{1}(t){\bf B}_{\rm e}(t)
-w_{2}(t) {\bf B}_{\rm e}^{-1}(t) \Bigr ]^{\prime}
:{\bf D}(t)
\nonumber\\
&& -\Bigl [ w_{1}(t){\bf B}_{\rm e}(t)
-w_{2}(t) {\bf B}_{\rm e}^{-1}(t) \Bigr ]^{\prime}
:{\bf M}(t)
\biggr \}+{\bf \Sigma}^{\prime}(t):{\bf D}(t)\geq 0,
\end{eqnarray}
where the functions $w_{m}(t)$ are given by
\begin{equation}
w_{m}(t)=\frac{\partial w}{\partial J_{{\rm e}m}}
\Bigl (J_{\rm e1}(t),J_{\rm e2}(t)\Bigr )
\qquad
(m=1,2).
\end{equation}
The dissipation inequality (26) is satisfied for an
arbitrary deformation program, provided that
the stress tensor reads
\begin{equation}
{\bf \Sigma}(t)=-P(t){\bf I}+2 N \Bigl (1-\beta(t)\Bigr )
\Bigl [ w_{1}(t){\bf B}_{\rm e}(t)
-w_{2}(t) {\bf B}_{\rm e}^{-1}(t) \Bigr ],
\end{equation}
and the tensor ${\bf M}(t)$ is given by
\begin{equation}
{\bf M}(t)=\alpha(t)\Bigl [ w_{1}(t){\bf B}_{\rm e}(t)
-w_{2}(t) {\bf B}_{\rm e}^{-1}(t) \Bigr ]^{\prime}.
\end{equation}
Here $P(t)$ is an unknown pressure, and $\alpha(t)$
is a non-negative scalar function.

It follows from Eqs. (12) and (29) that
\begin{eqnarray*}
{\bf M}\cdot{\bf B}_{\rm e}+{\bf B}_{\rm e}\cdot {\bf M}
& =& 2\alpha \Bigl [w_{1}\Bigl ({\bf B}_{\rm e}^{2}
-\frac{1}{3}J_{\rm e1}{\bf B}_{\rm e}\Bigr )
-w_{2}\Bigl ({\bf I}-\frac{1}{3} J_{\rm e2}{\bf B}_{\rm e}
\Bigr )\Bigr ].
\end{eqnarray*}
Substitution of this expression into Eq. (24) results
in the kinetic equation
\begin{eqnarray}
\frac{d{\bf B}_{\rm e}}{dt}(t)
&=& \Bigl (1-\gamma(t) \Bigr )
\Bigl [{\bf L}(t)\cdot {\bf B}_{\rm e}(t)
+{\bf B}_{\rm e}(t)\cdot {\bf L}^{\top}(t)\Bigr ]
\nonumber\\
&& +\Bigl (\gamma(t)-\beta(t)\Bigr )
\Bigl [ {\bf D}(t)\cdot {\bf B}_{\rm e}(t)
+{\bf B}_{\rm e}(t)\cdot {\bf D}(t)\Bigr ]
\nonumber\\
&&-2\alpha(t)\Bigl [w_{1}(t) {\bf B}_{\rm e}^{2}(t)
+\frac{1}{3}\Bigl (w_{2}(t) J_{\rm e2}(t)
-w_{1}(t) J_{\rm e1}(t) \Bigr ){\bf B}_{\rm e}(t)
-w_{2}(t){\bf I}\Bigr ].
\end{eqnarray}
Given a deformation program, Eqs. (28) and (30)
provide a set of constitutive equations that
involve four functions to be determined:
the strain energy per strand
$w(J_{\rm e1},J_{\rm e2})$, the function $\alpha(t)$
that characterizes the rate of internal dissipation,
and the functions $\beta(t)$ and $\gamma(t)$
that establish connections between the rate-of-strain
tensors and vorticity tensors for sliding of
junctions and macro-deformation, respectively.
It is worth noting that Eqs. (28) and (30) are applicable
for the description of the mechanical response of both
solid polymers and polymer fluids, because no assumptions
were introduced in their derivation regarding
the specific properties of a network of strands.

An important feature of polymer fluids is that their
time-dependent behavior reflects two different mechanisms
at the micro-level: (i) sliding of junctions between
chains with respect to their reference positions,
and (ii) slippage of chains with respect to entanglements.
Two approaches are conventionally employed to take
into account the latter process (which is not included
into the present model).

According to the first (the concept of transient networks
\cite{TE92}),
it is assumed that strands are not permanently connected
to their junctions, but detach from the junctions at random
instants, see Figure 1 in \cite{TE92}.
When an end of an active strand separates from a junction,
the strand is transformed into the dangling state,
where the stress totally relaxes.
When a free end of a dangling strand merges with a nearby
junction at random time, the stress-free state of
this strand coincides with the actual state of the network.

According to the other approach (the reptation theory
\cite{DE86}),
relaxation of stresses in chains occurs due to
(i) their curvilinear diffusion along the tubes formed
by surrounding macromolecules, and (ii) partial release
from the tubes, see Figures 6.1 and 6.4 in \cite{DE86}.
As a result of this motion, the reference state
of a strand differs from that of the vector
that connects its junction points.
This idea provides a physical ground for the pom--pom
model \cite{ML98}, where the average length of a strand
in its stress-free state is introduced as an independent
variable in the governing relations.

Despite evident merits of these concepts, they
share an important shortcoming:
these approaches lead to (i) a substantial
complication of the constitutive equations and
(ii) a noticeable increase in the number of
material functions and parameters.
To avoid an undesirable growth of experimental
constants, on the one hand, and to make the
constitutive equations thermodynamically consistent,
on the other, we postulate that the characteristic
rate of stress relaxation in strands (driven
by their curvilinear diffusion in tubes or by their
rearrangement in a transient network) substantially
exceeds the rate of sliding of junctions.
This implies that the number of strands per unit
volume of a network $N$ should be treated not as a
constant, but as a function of time whose values
are rapidly (compared to the characteristic time
of the deformation process) tuned to those
that are determined by the current state of the network
(an estimate for the relaxation time will be provided
in Section 9).
Our hypothesis is in accord with the double reptation
concept \cite{Clo88} that treat the response of
a network of macromolecules as that induced by
(i) rapid diffusive motion of chains in tubes
(which is disregarded in the present model),
and (ii) relatively slow relaxation of constrains
imposed by surrounding macromolecules on the motion
of a characteristic chain (which is accounted for
in terms of the sliding process).

The assumption regarding two relaxation processes
with different time scales implies that the
Clausius--Duhem inequality (26) remains valid
for relatively slow deformation processes with $N$
thought of as a function of time (according to the
Doi--Edwards theory \cite{DE86}, $N(t)$ is the
number of strands per unit volume where stresses
have not relaxed until instant $t$;
in terms of the pom--pom model \cite{ML98},
$N(t)$ is proportional to the square of the stretch
of a characteristic chain).
According to this approach, the kinetic equation
(30) remains unchanged, while the expression (28)
for the stress tensor reads
\begin{equation}
{\bf \Sigma}(t)=-P(t){\bf I}+2 N(t)
\Bigl (1-\beta(t)\Bigr )
\Bigl [ w_{1}(t){\bf B}_{\rm e}(t)
-w_{2}(t) {\bf B}_{\rm e}^{-1}(t) \Bigr ],
\end{equation}
where $N(t)$ is an adjustable function.

It is evident that the number of adjustable functions
in Eqs. (30) and (31) is too large to find them
with an acceptable level of accuracy by fitting
experimental data in conventional rheological tests.
To reduce this number, additional simplifications
are conventionally introduced.

{\bf 1.} Assuming strands to be Gaussian with the strain
energy density
\begin{equation}
w(J_{\rm e1},J_{\rm e2})=\frac{1}{2} \mu (J_{\rm e1}-3),
\end{equation}
where $\mu$ is the average rigidity of a strand,
setting $\alpha(t)=0$ and $\gamma(t)=0$,
and using Eq. (18), we arrive at the relations
\begin{eqnarray}
{\bf \Sigma}(t) &=& -P(t){\bf I}
+ N(t) \mu {\bf B}_{\rm e}(t),
\\
\frac{d{\bf B}_{\rm e}}{dt}(t) &=&
\Bigl (1-\beta(t)\Bigr )\Bigl [
{\bf D}(t)\cdot {\bf B}_{\rm e}(t)
+{\bf B}_{\rm e}(t)\cdot {\bf D}(t)\Bigr ]
+\Bigl [ {\bf \Omega}(t)\cdot {\bf B}_{\rm e}(t)
-{\bf B}_{\rm e}(t)\cdot {\bf \Omega}(t) \Bigr ].
\end{eqnarray}
We now introduce the effective stress by the formula
\begin{equation}
{\bf T}(t)=N(t) \mu {\bf B}_{\rm e}(t)-\Lambda(t){\bf I},
\end{equation}
where the function $\Lambda(t)$ will be specified later.
Differentiation of Eq. (35) with respect to time $t$ results
in
\[
\frac{d{\bf T}}{dt}=\frac{dN}{dt}(t)\mu {\bf B}_{\rm e}(t)
+N(t)\mu \frac{d{\bf B}_{\rm e}}{dt}(t)
-\frac{d\Lambda}{dt}(t){\bf I}.
\]
Substitution of Eqs. (34) and (35) into this equality
implies that
\begin{eqnarray}
\frac{d{\bf T}}{dt}(t) &=& \Bigl (1-\beta(t)\Bigr )
\Bigl [ {\bf D}(t)\cdot {\bf T}(t)
+{\bf T}(t)\cdot {\bf D}(t)\Bigr ]
+\Bigl [ {\bf \Omega}(t)\cdot {\bf T}(t)
-{\bf T}(t)\cdot {\bf \Omega}(t) \Bigr ]
\nonumber\\
&& +\frac{1}{N(t)}\frac{dN}{dt}(t){\bf T}(t)
+2\Bigl (1-\beta(t)\Bigr )\Lambda(t){\bf D}(t)
+\Bigl [ \frac{\Lambda(t)}{N(t)} \frac{dN}{dt}(t)
-\frac{d\Lambda}{dt}(t) \Bigr ]{\bf I}.
\end{eqnarray}
The function $\Lambda(t)$ is determined from the
condition that the last term on the right-hand side of
Eq. (36) vanishes,
\[
\Lambda(t)=C N(t),
\]
where $C$ is an arbitrary constant.
Setting
\[
\frac{1}{N(t)}\frac{dN}{dt}(t)=-\frac{1}{\tau(t)},
\]
where $\tau(t)$ is a new material function
(the characteristic time for rearrangement of tubes
according to the reptation theory), we find that
\[
N(t)=N(0)\exp \Bigl [-\int_{0}^{t} \tau(s) ds \Bigr ].
\]
Substitution of these expressions into Eq. (36)
results in the constitutive equation of
the Johnson--Segalman model \cite{JS77},
\begin{eqnarray}
\frac{d{\bf T}}{dt}(t) &=& \Bigl (1-\beta(t)\Bigr )
\Bigl [ {\bf D}(t)\cdot {\bf T}(t)
+{\bf T}(t)\cdot {\bf D}(t)\Bigr ]
+\Bigl [ {\bf \Omega}(t)\cdot {\bf T}(t)
-{\bf T}(t)\cdot {\bf \Omega}(t) \Bigr ]
\nonumber\\
&& +\frac{1}{\tau(t)} \Bigl [ 2\eta(t){\bf D}(t)
-{\bf T}(t)\Bigr ],
\end{eqnarray}
where the viscosity $\eta$ is given by
\[
\eta(t)=CN_{0}\Bigl (1-\beta(t)\Bigr )
\tau(t)\exp \Bigl [-\int_{0}^{t} \tau(s) ds \Bigr ].
\]

{\bf 2.} Adopting hypothesis (32) and setting
$\alpha(t)=0$ and $\gamma(t)=0$, we find that
\begin{eqnarray*}
{\bf \Sigma}(t) &=& -P(t){\bf I}+ N(t) \mu \Bigl (
1-\beta(t)\Bigr ) {\bf B}_{\rm e}(t),
\nonumber\\
{\bf B}_{\rm e}^{\triangle}(t) &=& -\beta(t)
\Bigl [{\bf D}(t)\cdot {\bf B}_{\rm e}(t)
+{\bf B}_{\rm e}(t)\cdot{\bf D}(t) \Bigr ],
\end{eqnarray*}
where
\[
{\bf B}_{\rm e}^{\triangle}(t)
=\frac{d{\bf B}_{\rm e}}{dt}(t)
-{\bf L}(t)\cdot {\bf B}_{\rm e}(t)
-{\bf B}_{\rm e}(t)\cdot {\bf L}^{\top}(t)
\]
is the Oldroyd contravariant derivative.
Introducing the effective stress ${\bf T}(t)$ by
Eq. (35), we obtain
\begin{eqnarray*}
{\bf T}^{\triangle}(t) &=& -\beta(t)
\Bigl [{\bf D}(t)\cdot {\bf T}(t)
+{\bf T}(t)\cdot{\bf D}(t) \Bigr ]
+\frac{1}{N(t)}\frac{dN}{dt}(t){\bf T}(t)
\nonumber\\
&& +2\Bigl (1 -\beta(t)\Bigr )\Lambda(t) {\bf D}(t)
+\Bigl [ \frac{\Lambda(t)}{N(t)} \frac{dN}{dt}(t)
-\frac{d\Lambda}{dt}(t) \Bigr ]{\bf I}.
\end{eqnarray*}
Equating the last term in the right-hand side of this
formula to zero, we arrive at the Phan-Thien--Tanner
constitutive model \cite{PTT77}
\begin{equation}
{\bf T}^{\triangle}(t) = -\beta(t)
\Bigl [{\bf D}(t)\cdot {\bf T}(t)
+{\bf T}(t)\cdot{\bf D}(t) \Bigr ]
+\frac{1}{\tau(t)} \Bigl [ 2\eta(t){\bf D}(t)
-{\bf T}(t)\Bigr ].
\end{equation}

{\bf 3.} Accepting formula Eq. (32) for the strain energy
of a Gaussian strand and assuming that $\beta(t)=0$ and
$\gamma(t)=0$, we find that the Cauchy stress tensor
${\bf \Sigma}(t)$ is given by Eq. (33),
where the function ${\bf B}_{\rm e}(t)$ is governed by
the differential equation
\[
{\bf B}_{\rm e}^{\triangle}(t)
= -\alpha(t) \mu \Bigl [{\bf B}_{\rm e}^{2}(t)
-\frac{1}{3} J_{\rm e1}(t) {\bf B}_{\rm e}(t)\Bigr ].
\]
Substitution of expression (35) for the effective stress
${\bf T}(t)$ into this equality implies that
\begin{eqnarray*}
{\bf T}^{\triangle}(t) &=& \Bigl [ \frac{1}{N(t)}
\frac{dN}{dt}(t)+2\frac{\alpha(t)\Lambda(t)}{N(t)}
-\frac{1}{3}\alpha \mu J_{\rm e1}(t)\Bigr ]{\bf T}(t)
-\frac{\alpha(t)}{N(t)}{\bf T}^{2}(t)
\nonumber\\
&& +2\Lambda(t){\bf D}(t)
+\Bigl [ \frac{\Lambda(t)}{N(t)}\frac{dN}{dt}(t)
-\frac{\alpha(t)}{N(t)}\Lambda^{2}(t)
-\frac{1}{3}\alpha(t)\mu \Lambda(t)J_{\rm e1}(t)
-\frac{d\Lambda}{dt}(t)\Bigr ]{\bf I}.
\end{eqnarray*}
Equating the last term of the right-hand side of this
equation to zero,
which results in the differential equation for
the function $\Lambda(t)$,
\[
\frac{d\Lambda}{dt}(t)=\Lambda(t)
\Bigl [ \frac{1}{N(t)}\frac{dN}{dt}(t)
-\frac{\alpha(t)}{N(t)}\Lambda(t)
-\frac{1}{3}\alpha(t)\mu J_{\rm e1}(t)\Bigr ],
\]
and introducing the new adjustable functions
\[
\tau(t)=-\Bigl [ \frac{1}{N(t)}
\frac{dN}{dt}(t)+2\frac{\alpha(t)\Lambda(t)}{N(t)}
-\frac{1}{3}\alpha \mu J_{\rm e1}(t)\Bigr ],
\qquad
\eta(t)=\Lambda(t)\tau(t),
\]
we arrive at the formula
\begin{equation}
{\bf T}^{\triangle}(t)+\frac{\alpha(t)}{N(t)}{\bf T}^{2}(t)
= \frac{1}{\tau(t)} \Bigl [
2\eta(t){\bf D}(t)-{\bf T}(t)\Bigr ],
\end{equation}
which coincides with the Giesekus constitutive
equation \cite{Gie82}.

It is worth noting that the material functions $\tau(t)$
and $\eta(t)$ in Eqs. (37) to (39) are not arbitrary,
but they obey some additional restrictions.
This is caused by the fact that we consider relaxation of
stresses in strands driven by their diffusion inside tubes
as a rapid process compared to rearrangement of tubes
(treated as sliding of junctions).
An advantage of our approach is that it operates
with the Cauchy deformation tensors conventionally
employed in continuum mechanics (no configurational
tensors are introduced) and is based on the classical
multiplicative decomposition of the deformation
gradient (unlike constitutive equations that explicitly
use a slip tensor).

{\bf 4.} Finally, under the assumptions that (i)
$w$ is a symmetric function of its arguments,
which implies that
\[
w_{1}(t)=w_{2}(t)=\bar{w}(t),
\]
and (ii) the functions $\beta(t)$ and $\gamma(t)$
vanish,
Eqs. (28) and (31) are transformed into the Leonov
constitutive model \cite{SL95}
\begin{eqnarray*}
{\bf \Sigma}(t) &=& -P(t){\bf I}+2N(t) \bar{w}(t)\Bigl (
{\bf B}_{\rm e}(t)-{\bf B}_{\rm e}^{-1}(t)\Bigr ),
\nonumber\\
{\bf B}_{\rm e}^{\triangle}(t)
&=& -2\alpha(t) \bar{w}(t)
\Bigl [ {\bf B}_{\rm e}^{2}(t)
+\frac{1}{3}\Bigl (J_{\rm e2}(t)-J_{\rm e1}(t)
\Bigr ){\bf B}_{\rm e}(t)-{\bf I}\Bigr ].
\end{eqnarray*}

Unlike the above approaches, this study focuses
on the case when the functions $\beta(t)$ and
$\gamma(t)$ coincide,
\begin{equation}
\gamma(t)=\beta(t)=b(t).
\end{equation}
This allows us to omit the second term on the
right-hand side of Eq. (30), which may result
in stress oscillations at simple shear that have
no physical meaning.
From the kinematic standpoint, condition (40)
means that the velocity gradients for sliding
of junctions and macro-deformation are connected
by the linear equation, see Eqs. (18) and (23),
\[
{\bf L}_{\rm s}(t)= b(t){\bf L}(t)+{\bf M}(t),
\]
where ${\bf M}(t)$ is a symmetric traceless tensor
given by Eq. (29).
Substitution of expression (40) into Eq. (30) implies
the differential equation
\begin{eqnarray}
\frac{d{\bf B}_{\rm e}}{dt}(t)
&=& \Bigl (1-b(t) \Bigr )
\Bigl [{\bf L}(t)\cdot {\bf B}_{\rm e}(t)
+{\bf B}_{\rm e}(t)\cdot {\bf L}^{\top}(t)\Bigr ]
\nonumber\\
&&-2\alpha(t)\Bigl [w_{1}(t) {\bf B}_{\rm e}^{2}(t)
+\frac{1}{3}\Bigl (w_{2}(t) J_{\rm e2}(t)
-w_{1}(t) J_{\rm e1}(t) \Bigr ){\bf B}_{\rm e}(t)
-w_{2}(t){\bf I}\Bigr ].
\end{eqnarray}
For a network of Gaussian strands, see Eq. (32),
the stress--strain relations (31) and (41) read
\begin{eqnarray}
\hspace*{-5 mm}{\bf \Sigma}(t) &=& -P(t){\bf I}+ G(t)
\Bigl (1-b(t)\Bigr ) {\bf B}_{\rm e}(t),
\nonumber\\
\hspace*{-5 mm}\frac{d{\bf B}_{\rm e}}{dt}(t)
&=& \Bigl (1-b(t) \Bigr )
\Bigl [{\bf L}(t)\cdot {\bf B}_{\rm e}(t)
+{\bf B}_{\rm e}(t)\cdot {\bf L}^{\top}(t)\Bigr ]
-a(t) \Bigl [{\bf B}_{\rm e}^{2}(t)
-\frac{1}{3} J_{\rm e1}(t) {\bf B}_{\rm e}(t) \Bigr ]
\end{eqnarray}
with
\[
a(t)=\alpha(t)\mu,
\qquad
G(t)=\mu N(t).
\]
Our aim now is to specify the dependencies of $a$, $b$
and $G$ in Eqs. (42) on parameters that characterize
macro-deformation of a polymer fluid.

\section{Adjustable functions}

We introduce the rate intensity $D_{\rm i}$ by the
conventional formula
$D_{\rm i}=\Bigl (2 {\bf D}:{\bf D}\Bigr )^{\frac{1}{2}}$
and adopt the following equation for the elastic
modulus $G$:
\begin{equation}
\frac{\ln G}{\ln G_{0}}=\exp \Bigl (K_{G}D_{\rm i}^{\nu_{G}}
\Bigr ),
\end{equation}
where $G_{0}$, $K_{G}$ and $\nu_{G}$ are non-negative
material constants.
According to Eq. (43), $G_{0}$ is the plateau modulus
at small strain rates, whereas $K_{G}$ and $\nu_{G}$
characterize a strong increase in $G$ with strain-rate
intensity.

The parameter $b$ is assumed to obey the power law,
\begin{equation}
b=K_{b}D_{\rm i}^{\nu_{b}}
\end{equation}
with non-negative adjustable parameters $K_{b}$
and $\nu_{b}$.
Formula (44) means that no sliding of junctions occurs
at very small strain rates, and the rate of sliding
grows with the rate of macro-strain.
Equations (43) and (44) are applicable within a limited
range of strain rates, as the modulus $G$ and
the parameter $b$ should reach some limiting values
at sufficiently high strain rates.

The coefficient $a$ is split into the product of two
functions
\begin{equation}
a=\zeta A,
\end{equation}
where the function $\zeta$ characterizes the effect
of strain and the function $A$ determines the effect
of strain rate on the rate of energy dissipation
[described by means of the coefficient $a$ that is
proportional to $\alpha$ in Eqs. (26) and (29)].

The dependence of the parameter $A$ on strain-rate
intensity $D_{\rm i}$ is determined by the power-law
similar to Eq. (44),
\begin{equation}
A=A_{0}+K_{A}D_{\rm i}^{\nu_{A}},
\end{equation}
where $A_{0}$, $K_{A}$ and $\nu_{A}$ are non-negative
coefficients.
Eq. (46) means that $A$ monotonically increases with
strain rate from its plateau value $A_{0}$.
Unlike Eq. (44) for the function $b(D_{\rm i})$,
we do not presume $A$ to vanish at small strain rates;
a reason for this claim will be explained later
in Section 9.

It seems natural to suppose that for a developed flow
with large strains,
the motion of junctions at the micro-level is affine
with macro-deformation.
It follows from Eqs. (23) and (40) that this hypothesis
is tantamount to the assertion that the coefficient $\alpha$
in Eq. (29) vanishes.
This conclusion implies that $\zeta$ should approach zero
at large deformations (that is at large strain energies).
To approximate such a dependence, the exponential
function is chosen,
\begin{equation}
\zeta=\exp \Bigl [-r (J_{\rm e1}-3)\Bigr ],
\end{equation}
where $r\geq 0$ is a constant.
Our choice of the term $J_{\rm e1}-3$ in
Eq. (47) may be justified by the fact that this
expression is proportional to the strain energy of
a Gaussian strand (32).
Combining Eqs. (45) to (47), we arrive at the formula
\begin{equation}
a=\Bigl (A_{0}+K_{A}D_{\rm i}^{\nu_{A}}\Bigr )
\exp \Bigl [-r (J_{\rm e1}-3)\Bigr ].
\end{equation}
Formally, Eqs. (42) to (44) and (48) involve 9 adjustable
parameters: $A_{0}$, $G_{0}$, $K_{A}$, $K_{b}$,
$K_{G}$, $\nu_{A}$, $\nu_{b}$, $\nu_{G}$ and $r$.
It should be noted, however, that any set of experimental
data at an arbitrary deformation with a constant strain-rate
intensity $D_{\rm i}$ is determined by 4 constants only:
$A$, $b$, $G$ and $r$.
The other quantities can be found by fitting
the functions $A(D_{\rm i})$, $b(D_{\rm i})$
and $G(D_{\rm i})$ by Eqs. (43), (44) and (46).

Our aim now is to apply constitutive equations (42)
for the analysis of uniaxial extension and simple shear
of a polymer fluid.

\section{Uniaxial extension}

Uniaxial extension of an incompressible medium
is described by the formulas
\begin{equation}
x_{1}=k(t)X_{1},
\qquad
x_{2}=k^{-\frac{1}{2}}(t)X_{2},
\qquad
x_{3}=k^{-\frac{1}{2}}(t)X_{3},
\end{equation}
where $\{ X_{i} \}$ are Cartesian coordinates
in the initial state,
$\{ x_{i} \}$ are Cartesian coordinates in
the deformed state,
and $k=k(t)$ stands for elongation.
We suppose that transition from the initial
state to the current reference state that determines
the sliding process is described by the equations
similar to Eqs. (49),
\begin{equation}
\xi_{1}=\kappa(t)X_{1},
\qquad
\xi_{2}=\kappa^{-\frac{1}{2}}(t)X_{2},
\qquad
\xi_{3}=\kappa^{-\frac{1}{2}}(t)X_{3},
\end{equation}
where $\{ \xi_{i} \}$ are Cartesian coordinates
in the intermediate state, and $\kappa=\kappa(t)$
is a function to be found.
According to Eqs. (49) and (50),
the deformation gradients ${\bf F}$ and ${\bf F}_{\rm s}$
are given by
\begin{equation}
{\bf F}=k{\bf e}_{1}{\bf e}_{1}
+k^{-\frac{1}{2}}({\bf e}_{2}{\bf e}_{2}
+{\bf e}_{3}{\bf e}_{3}),
\qquad
{\bf F}_{\rm s}=\kappa{\bf e}_{1}{\bf e}_{1}
+\kappa^{-\frac{1}{2}}({\bf e}_{2}{\bf e}_{2}
+{\bf e}_{3}{\bf e}_{3}),
\end{equation}
where ${\bf e}_{i}$ are unit vectors of the frame
$\{ X_{i} \}$.
It follows from Eqs. (1), (8) and (51) that
\[
{\bf B}_{\rm e}=\Bigl (\frac{k}{\kappa}\Bigr )^{2}
{\bf e}_{1}{\bf e}_{1}
+\frac{\kappa}{k}({\bf e}_{2}{\bf e}_{2}
+{\bf e}_{3}{\bf e}_{3})
\]
and
\[
J_{\rm e1}=\Bigl (\frac{k}{\kappa}\Bigr )^{2}
+2\frac{\kappa}{k}.
\]
The velocity gradient ${\bf L}$, the rate-of-strain tensor
${\bf D}$, and the strain-rate intensity $D_{\rm i}$ read
\begin{equation}
{\bf L}={\bf D}=\frac{\dot{k}}{k}\Bigl [
{\bf e}_{1}{\bf e}_{1}
-\frac{1}{2}({\bf e}_{2}{\bf e}_{2}
+{\bf e}_{3}{\bf e}_{3})\Bigr ],
\qquad
D_{\rm i}=\frac{|\dot{k}|}{k}\sqrt{3},
\end{equation}
where $\dot{k}=dk/dt$.
Substitution of these expressions into Eqs. (42)
results in the differential equation for the
function $\kappa(t)$,
\begin{equation}
\frac{\dot{\kappa}}{\kappa}
=b\frac{\dot{k}}{k}+ \frac{a}{3}
\Bigl [ \Bigl (\frac{k}{\kappa}\Bigr )^{2}
-\frac{\kappa}{k}\Bigr ],
\qquad
\kappa(0)=1.
\end{equation}
The Cauchy stress tensor ${\bf \Sigma}$ is given by
\[
{\bf \Sigma}=\Sigma_{1}{\bf e}_{1}{\bf e}_{1}
+\Sigma_{2}({\bf e}_{2}{\bf e}_{2}
+{\bf e}_{3}{\bf e}_{3}),
\]
where the stress difference
$\Delta\Sigma=\Sigma_{1}-\Sigma_{2}$ reads
\begin{equation}
\Delta \Sigma=G(1-b)\Bigl [
\Bigl (\frac{k}{\kappa}\Bigr )^{2}
-\frac{\kappa}{k}\Bigr ].
\end{equation}
Given a deformation program $k(t)$, Eqs. (53) and (54)
together with the phenomenological equations
(43), (44) and (48) determine the stress difference
$\Delta \Sigma$ as a function of time.

For uniaxial extension with a constant rate of Hencky
strain
\[
\frac{\dot{k}}{k}=\dot{\epsilon},
\]
it is natural to search a solution of Eq. (53) in the form
\begin{equation}
\kappa(t)=z(t)\exp(\dot{\epsilon}t),
\end{equation}
where $z(t)$ is a function to be found.
Substituting expression (55) into Eq. (53)
and using Eq. (48), we find that
\begin{equation}
\dot{z}+(1-b)\dot{\epsilon} z=\frac{A}{3z}(1-z^{3})
\exp\Bigl (-r\frac{1-3z^{2}+2z^{3}}{z^{2}}\Bigr ),
\qquad
z(0)=1,
\end{equation}
where $A$, $b$ and $G$ depends on $\dot{\epsilon}$.
It follows from Eq. (56) that when $t\to\infty$,
the function $z(t)$ approaches its limiting value $z_{0}$,
which is determined by the nonlinear equation
\begin{equation}
\frac{1-z_{0}^{3}}{z_{0}^{2}}
\exp\Bigl (-r\frac{1-3z_{0}^{2}+2z_{0}^{3}}{z_{0}^{2}}\Bigr )
=\frac{3(1-b)\dot{\epsilon}}{A}.
\end{equation}
According to Eqs. (54) and (55), the transient
elongational viscosity
\begin{equation}
\eta_{\rm e}^{+}=\frac{\Delta \Sigma}{\dot{\epsilon}}
\end{equation}
is given by
\[
\eta_{\rm e}^{+}=G(1-b)\frac{1-z^{3}}{\dot{\epsilon}z^{2}}.
\]
Equation Eq. (57) implies that when $t\to\infty$,
the transient viscosity $\eta_{\rm e}^{+}$ tends to
its limiting value
\[
\eta_{\rm e}=\frac{3G}{A}(1-b)^{2}\exp
\Bigl (r\frac{1-3z_{0}^{2}+2z_{0}^{3}}{z_{0}^{2}}\Bigr ).
\]
This formula together with Eq. (57) determine the steady
elongational viscosity $\eta_{\rm e}$ as a function
of the Hencky strain rate $\dot{\epsilon}$.
When $r=0$ (no effect of strain on the coefficient $a$),
we find that
\begin{equation}
\eta_{\rm e}=\frac{3G}{A}(1-b)^{2}.
\end{equation}
According to Eqs. (44) and (48), Eq. (59) describes
a pronounced decrease in the steady elongational
viscosity at large strain rates observed in numerous
experiments.

\section{Simple shear}

Simple shear of an incompressible medium is described
by the equations
\begin{equation}
x_{1}=X_{1}+k(t)X_{2},
\qquad
x_{2}=X_{2},
\qquad
x_{3}=X_{3},
\end{equation}
where $\{ X_{i}\}$ and $\{ x_{i} \}$ are Cartesian
coordinates in the initial and actual states,
respectively, and $k(t)$ stands for shear.
We describe transition from the initial to the
intermediate configuration that characterizes sliding
of junctions as a superposition of
simple shear and three-dimensional extension,
\begin{equation}
\xi_{1}=\lambda_{1}(t)X_{1}+\kappa(t)X_{2},
\qquad
\xi_{2}=\lambda_{2}(t)X_{2},
\qquad
\xi_{3}=\lambda_{3}(t)X_{3},
\end{equation}
where $\{ \xi_{i} \}$ are Cartesian coordinates
in the stress-free state, and $\lambda_{i}(t)$
and $\kappa(t)$ are functions to be found.
The functions $\lambda_{i}(t)$ obey the
incompressibility condition
\[
\lambda_{1}\lambda_{2}\lambda_{3}=1.
\]
It follows from Eqs. (60) and (61) that the
deformation gradients ${\bf F}$ and ${\bf F}_{\rm s}$
read
\begin{eqnarray*}
{\bf F} &=& {\bf e}_{1}{\bf e}_{1}
+{\bf e}_{2}{\bf e}_{2}
+{\bf e}_{3}{\bf e}_{3}
+k{\bf e}_{1}{\bf e}_{2},
\nonumber\\
{\bf F}_{\rm s} &=& \lambda_{1}{\bf e}_{1}{\bf e}_{1}
+\lambda_{2}{\bf e}_{2}{\bf e}_{2}
+\lambda_{3}{\bf e}_{3}{\bf e}_{3}
+\kappa{\bf e}_{1}{\bf e}_{2}.
\end{eqnarray*}
Substitution of these expressions into Eq. (1)
implies that
\begin{equation}
{\bf F}_{\rm e} = p_{1}{\bf e}_{1}{\bf e}_{1}
+p_{2}{\bf e}_{2}{\bf e}_{2}
+p_{3}{\bf e}_{3}{\bf e}_{3}
+\phi{\bf e}_{1}{\bf e}_{2},
\end{equation}
where
\[
p_{1}=\lambda_{2}\lambda_{3},
\qquad
p_{2}=\lambda_{1}\lambda_{3},
\qquad
p_{3}=\lambda_{1}\lambda_{2},
\qquad
\phi=\lambda_{3}(\lambda_{1}k-\kappa).
\]
In the new notation, the incompressibility condition
is given by
\begin{equation}
p_{1}p_{2}p_{3}=1.
\end{equation}
It follows from Eqs. (8) and (62) that the left
Cauchy--Green tensor for elastic deformation is
determined as
\begin{equation}
{\bf B}_{\rm e}
= (p_{1}^{2}+\phi^{2}){\bf e}_{1}{\bf e}_{1}
+p_{2}^{2}{\bf e}_{2}{\bf e}_{2}
+p_{3}^{2}{\bf e}_{3}{\bf e}_{3}
+p_{2}\phi({\bf e}_{1}{\bf e}_{2}
+{\bf e}_{2}{\bf e}_{1}).
\end{equation}
The velocity gradient ${\bf L}$, the rate-of-strain
tensor ${\bf D}$, and the strain-rate intensity $D_{\rm i}$
read
\begin{equation}
{\bf L}=\dot{k}{\bf e}_{1}{\bf e}_{2},
\qquad
{\bf D}=\frac{1}{2}\dot{k}
({\bf e}_{1}{\bf e}_{2}
+{\bf e}_{2}{\bf e}_{1}),
\qquad
D_{\rm i}=\dot{k}.
\end{equation}
We substitute expressions (64) and (65) into the second
equation in Eqs. (42).
Omitting simple but tedious algebra, we arrive at the
following equations:
\begin{eqnarray}
\dot{p}_{1} &=& -\frac{a}{6} p_{1}
(2p_{1}^{2}-p_{2}^{2}-p_{3}^{2}-\phi^{2}),
\qquad
p_{1}(0)=1,
\nonumber\\
\dot{p}_{2} &=& -\frac{a}{6} p_{2}
(-p_{1}^{2}+2p_{2}^{2}-p_{3}^{2}+2\phi^{2}),
\qquad
p_{2}(0)=1,
\nonumber\\
\dot{p}_{3} &=& -\frac{a}{6} p_{1}
(-p_{1}^{2}-p_{2}^{2}+2p_{3}^{2}-\phi^{2}),
\qquad
p_{3}(0)=1,
\nonumber\\
\dot{\phi} &=& (1-b)\dot{k}p_{2}
-\frac{a}{6} \phi
(5p_{1}^{2}+2p_{2}^{2}-p_{3}^{2}+2\phi^{2}),
\qquad
\phi(0)=0.
\end{eqnarray}
It is easy to check that Eqs. (66) imply the
incompressibility condition (63) for an
arbitrary deformation program $k(t)$.
Combining Eqs. (42) and (64), we obtain
\[
{\bf \Sigma}= \Sigma_{11}{\bf e}_{1}{\bf e}_{1}
+\Sigma_{22}{\bf e}_{2}{\bf e}_{2}
+\Sigma_{33}{\bf e}_{3}{\bf e}_{3}
+\Sigma_{12}\phi({\bf e}_{1}{\bf e}_{2}
+{\bf e}_{2}{\bf e}_{1}),
\]
where
\begin{eqnarray}
\Sigma_{11} &=& -P+G(1-b)(p_{1}^{2}+\phi^{2}),
\nonumber\\
\Sigma_{22} &=& -P+G(1-b)p_{2}^{2},
\nonumber\\
\Sigma_{33} &=& -P+G(1-b)p_{3}^{2},
\nonumber\\
\Sigma_{12} &=& G(1-b)p_{2}\phi.
\end{eqnarray}
It follows from these equations that the first
normal stress difference $N_{1}^{+}=\Sigma_{11}-\Sigma_{22}$
and the second normal stress difference
$N_{2}^{+}=\Sigma_{22}-\Sigma_{33}$ read
\begin{equation}
N_{1}^{+}=G(1-b)(p_{1}^{2}-p_{2}^{2}+\phi^{2}),
\qquad
N_{2}^{+}=G(1-b)(p_{2}^{2}-p_{3}^{2}).
\end{equation}
Equations (66) to (68) describe the time-dependent
response of a polymer fluid under shear with an
arbitrary deformation program $k(t)$.
Our aim now is to analyze these equations for
shearing with a constant strain rate $\dot{k}$ and to
study their steady-state solutions at $t\to\infty$.
Assuming that $p_{i}(t)\to p_{i0}$ and $\phi(t)\to\phi_{0}$
as $t\to\infty$, we find from Eqs. (66) that the limiting
values $p_{i0}$ and $\phi_{0}$ satisfy the algebraic
equations
\begin{eqnarray}
2p_{10}^{2}-p_{20}^{2}-p_{30}^{2}-\phi_{0}^{2} &=& 0,
\nonumber\\
-p_{10}^{2}+2p_{20}^{2}-p_{30}^{2}+2\phi_{0}^{2} &=& 0,
\nonumber\\
-p_{10}^{2}-p_{20}^{2}+2p_{30}^{2}-\phi_{0}^{2} &=& 0.
\end{eqnarray}
To derive these equations, the incompressibility
condition (63) is employed which excludes the
case when any of $p_{i0}$ vanishes.
Subtracting the third equality in Eqs. (69) from
the first, we find that
\begin{equation}
p_{30}^{2}=p_{10}^{2}.
\end{equation}
Substitution of expression (70) into the first equality
in Eqs. (69) results in
\begin{equation}
p_{10}^{2}=p_{20}^{2}+\phi_{0}^{2}.
\end{equation}
Combining Eqs. (63), (70) and (71), we obtain
\begin{equation}
p_{20}(p_{20}^{2}+\phi_{0}^{2})=1.
\end{equation}
It follows from Eqs. (66) that $\phi_{0}$
satisfies the equation
\[
(1-b)\dot{k}p_{20}=
\frac{1}{6}\zeta_{0}A \phi_{0}
(5p_{10}^{2}+2p_{20}^{2}-p_{30}^{2}+2\phi_{0}^{2}),
\]
where, according to Eqs. (46) and (64),
\[
\zeta_{0}=\exp \Bigl [-r\Bigl (p_{10}^{2}
+p_{20}^{2}+p_{30}^{2}+\phi_{0}^{2}-3\Bigr )\Bigr ].
\]
Substitution of expressions (70) and (71) into these
equations implies that
\[
\phi_{0}(p_{20}^{2}+\phi_{0}^{2})
=\frac{(1-b)\dot{k}}{\zeta_{0}A} p_{20},
\qquad
\zeta_{0}=\exp \Bigl [-3r\Bigl (p_{20}^{2}+\phi_{0}^{2}
-1\Bigr )\Bigr ].
\]
Excluding the sum $p_{20}^{2}+\phi_{0}^{2}$ by means of
Eq. (72), we find that
\begin{equation}
\phi_{0}=\frac{(1-b)\dot{k}}{A}p_{20}^{2}\exp \Bigl [
3r(p_{20}^{-1}-1)\Bigr ].
\end{equation}
Finally, substitution of expression (73) into Eq. (72)
results in the transcendental equation for $p_{20}$,
\begin{equation}
p_{20}^{3}\biggl [1+\frac{(1-b)^{2}\dot{k}^{2}}{A^{2}}
p_{20}^{2}\exp \Bigl (6r (p_{20}^{-1}-1)\Bigr )
\biggr ]=1.
\end{equation}
It follows from Eqs. (67) that the shear stress $\Sigma_{12}$
tends to its limiting value
\[
\Sigma_{12\;0}=G(1-b)p_{20}\phi_{0}
\]
when $t$ approaches infinity.
Substituting expression (73) into this formula
and introducing the steady shear viscosity
\[
\eta_{\rm s}=\frac{\Sigma_{12\;0}}{\dot{k}},
\]
we obtain
\begin{equation}
\eta_{\rm s}=\frac{G}{A}(1-b)^{2} p_{20}^{3}
\exp \Bigl [3r (p_{20}^{-1}-1)\Bigr ].
\end{equation}
In the case when $r=0$, which is equivalent to the
neglect of the influence of strain on the coefficient
$a$, Eq. (75) reads
\begin{equation}
\eta_{\rm s}=\frac{G}{A}(1-b)^{2} p_{20}^{3}.
\end{equation}
Comparing Eqs. (59) and (76) and introducing the
Trouton ratio
\[
{\rm Tr}=\frac{\eta_{\rm e}}{\eta_{\rm s}},
\]
we obtain
\[
{\rm Tr}=3p_{20}^{-3}.
\]
Combining this formula with Eq. (74) (where we set $r=0$)
and bearing in mind Eq. (65), we arrive at the equation
for the Trouton ratio
\[
{\rm Tr}=3\biggl \{1 +\Bigl [
\frac{3^{\frac{1}{3}} (1-b)D_{\rm i}}{A}\Bigr ]^{2}
{\rm Tr}^{-\frac{2}{3}}\biggr \}.
\]
This formula implies that under the condition
$\nu_{G}<1$, the Trouton ratio equals 3
at very small strain rates, and it increases
with the strain-rate intensity, at least
at relatively small $D_{\rm i}$,
in agreement with the experimental data reported in
\cite{BMT96}.

We return now to the general case $r\geq 0$
and calculate the limiting values $N_{1}$ and $N_{2}$
of the normal stress differences when $t\to\infty$.
It follows from Eqs. (68), (70) and (71) that
\[
N_{1}=2G(1-b)\phi_{0}^{2},
\qquad
N_{2}=-G(1-b)\phi_{0}^{2}.
\]
Introducing the steady normal stress functions
$\Psi_{i}=N_{i}\dot{k}^{-2}$
and using Eq. (73), we find that
\begin{equation}
\Psi_{1}=\frac{2G}{A^{2}}(1-b)^{3} p_{20}^{4}
\exp \Bigl [6r(p_{20}^{-1}-1)\Bigr ],
\qquad
\Psi_{2}=-\frac{1}{2}\Psi_{1}.
\end{equation}
Equations (74), (75) and (77) together with the
phenomenological relations (43), (44) and (46)
determine the steady shear viscosity and normal
stress functions for an arbitrary shear rate $\dot{k}$.
Equation (77) correctly predicts that the second
normal stress coefficient $\Psi_{2}$ is negative
and proportional to the first normal stress
function $\Psi_{1}$.
It is conventionally assumed for polymer fluids
that the coefficient of proportionality
between $\Psi_{1}$ and $\Psi_{2}$ should be of order
of 0.1 to 0.3 \cite{BMT96}, but the ratio
$|\Psi_{2}|/\Psi_{1}$ is strongly affected by
experimental conditions, and its precise value
remains unknown \cite{Sch02}.
Recent experiments on suspension in viscous fluids
show that this ratio is located in the interval
between 0.1 and 0.7 with the most probable value
of 0.5 for highly viscous liquids \cite{MGM02}.
The latter result is in excellent agreement with Eqs. (77).

\section{Fitting of observations}

Our aim now is to demonstrate that constitutive
equations (42) can correctly approximate observations
in start-up shear tests with various strain rates
and to show that the adjustable parameters
change consistently with shear rate
and characteristics of molecular weight distribution.
For this purpose, we focus on three sets of
experimental data for solutions of polystyrene in
tricresyl phosphate.
For a detailed description of the experimental procedure
and the material properties, we refer to the original
studies \cite{OLM98,OII00,OIU00a}.
The main parameters of the solutions (abbreviated here
as PS1, PS2 and PS3) are collected in Table 1.
The polydispersity index of PS2 and PS3 are not provided,
but it is mentioned that they have ``sharp molecular
weight distributions" \cite{OII00,OIU00a}.
Based on these observations, we treat the solutions
as monodisperse and concentrate on the effects of
mass-average molecular weight $M_{\rm w}$ and
the number of entanglements per chain $M_{\rm w}/M_{\rm e}$
($M_{\rm e}$ stands for the average molecular
weight between entanglements) on material constants.

We begin with fitting the observations on PS1
for the shear stress $\Sigma_{12}$ and the first normal
stress difference $N_{1}^{+}$ depicted in Figures 1
and 2.
These figures show that the functions $\Sigma_{12}(t)$
and $N_{1}^{+}(t)$ increase monotonically with time
at small shear rates $\dot{k}$ and
demonstrate pronounced stress overshoots
at relatively large $\dot{k}$.
As the experimental data were reported in Figure 4
of \cite{OLM98} with the use of the regular time-scale,
we follows the same approach.
The other sets of data (see Figures 6 to 8 below)
are presented in the logarithmic time-scale ($\log=\log_{10}$)
that makes more transparent the characteristic features
of stress overshoot.

First, we approximate the experimental dependencies
$\Sigma_{12}(t)$ and $N_{1}^{+}(t)$ measured at
the highest strain rate $\dot{k}_{\max}=4.0$ s$^{-1}$,
by using four constants, $A$, $b$, $G$ and $r$
[the coefficient $a$ in Eqs. (66) is calculated
from Eq. (45), where $\zeta$ is given by Eq. (47)].
To find the coefficients $A$, $b$, $G$ and $r$,
we fix some intervals $[0,A_{\max}]$, $[0,b_{\max}]$,
$[0,G_{\max}]$ and $[0,r_{\max}]$,
where the ``best-fit" parameters $A$, $b$, $G$
and $r$ are assumed to be located,
and divide these intervals into $J$ subintervals by
the points $A^{(i)}=i\Delta A$, $b^{(j)}=j\Delta b$,
$G^{(k)}=k\Delta G$ and $r^{(l)}=l\Delta r$
($i,j,k,l=1,\ldots,J-1$) with
$\Delta A=A_{\max}/J$, $\Delta b=b_{\max}/J$,
$\Delta G=G_{\max}/J$ and $\Delta r=r_{\max}/J$.
For any set $\{ A^{(i)}, b^{(j)}, G^{(k)}, r^{(l)} \}$,
Eqs. (66) are integrated numerically by the Runge--Kutta
method with the time-step $\Delta t=1.0\cdot 10^{-3}$
s$^{-1}$, and the quantities $\Sigma_{12}$ and $N_{1}^{+}$
are calculated from Eqs. (67) and (68).
The ``best-fit" parameters $A$, $b$, $G$ and $r$ are
determined from the condition of minimum of the
function
\begin{equation}
R=\sum_{t_{m}} \biggl \{ \Bigl [
\Sigma_{12}^{\rm exp}(t_{m})
-\Sigma_{12}^{\rm num}(t_{m})\Bigr ]^{2}
+ \delta \Bigl [ N_{1}^{\rm exp}(t_{m})
-N_{1}^{\rm num}(t_{m}) \Bigr ]^{2} \biggr \}
\end{equation}
on the set $ \{ A^{(i)}, b^{(j)}, G^{(k)}, r^{(l)}\;\;
(i,j,k,l=1,\ldots, J-1)\}$.
Here the coefficient $\delta$ characterizes a ``weight"
of the second term compared to the first one,
the sum is calculated over all experimental points
$t_{m}$ depicted in Figures 1 and 2,
$\Sigma_{12}^{\rm exp}$ and $N_{1}^{\rm exp}$
are the shear stress and the first normal stress
difference measured in a test,
and $\Sigma_{12}^{\rm num}$ and $N_{1}^{\rm num}$
are given by Eqs. (67) and (68).
In the numerical analysis, we set $J=10$
and $\delta=0.05$.
After finding the ``best-fit" values $A^{(i)}$, $b^{(j)}$,
$G^{(k)}$ and $r^{(l)}$, this procedure is repeated twice
for the new intervals $[ A^{(i-1)}, A^{(i+1)}]$,
$[ b^{(j-1)}, b^{(j+1)}]$, $[ G^{(k-1)}, G^{(k+1)}]$
and $[ r^{l-1)}, r^{(l+1)} ]$,
to ensure an acceptable accuracy of fitting.

After finding the ``best-fit" value of $r$ by matching
the data obtained at $\dot{k}_{\max}$, we fix
this parameter and approximate the observations
at other shear rates with the help of only three
material constants, $A$, $b$ and $G$.
Each pair of curves (for $\Sigma_{12}$ and $N_{1}^{+}$)
measured at a given shear rate $\dot{k}$ is
matched separately.

Figures 1 and 2 demonstrate good agreement between
the experimental data and the results of numerical
simulation.
The model correctly predicts that the shear stress
$\Sigma_{12}$ reaches its maximum before the first
normal stress difference $N_{1}^{+}$ approaches its
highest value.
It is worth mentioning that the quality of fitting
observations in Figures 1 and 2 is substantially better
than that reported for other constitutive models
in \cite{PL01} (underprediction of the shear stress)
and in \cite{HSJ99} (overprediction of $\Sigma_{12}$).

The quantities $A$, $b$ and $G$ are plotted
in Figures 3 to 5 versus shear rate $\dot{k}$.
The function $A(\dot{k})$ in Figure 3 is
approximated by the simplified version of Eq. (46)
with $A_{0}=0$.
Bearing in mind expression (65) for $D_{\rm i}$,
we present Eqs. (44) and (46) in the form
\begin{equation}
\log A=\log K_{A}+\nu_{A}\log \dot{k},
\qquad
\log b=\log K_{b}+\nu_{b}\log\dot{k}.
\end{equation}
The coefficients $K_{A}$, $K_{b}$, $\nu_{A}$ and
$\nu_{b}$ in Eqs. (79) are determined by
the least-squares technique.
The function $G(\dot{k})$ in Figure 5 is fitted
by Eq. (43), where the coefficients $G_{0}$,
$K_{G}$ and $\nu_{G}$ are found by the nonlinear
regression algorithm.
These quantities are listed in Table 2.
Figures 3 to 5 show reasonable agreement between
the experimental data and their approximations
by Eqs. (43) and (79).

We proceed with fitting observations for polystyrene
solutions PS2 and PS3, where we confine ourselves
to matching experimental data for the shear stress
$\Sigma_{12}$ only.
The same algorithm of fitting is employed as for PS1,
with the only exception that the last term in Eq. (78)
is omitted ($\delta=0$).

The experimental dependencies $\Sigma_{12}(t)$
measured at various strain rates $\dot{k}$
are plotted  together with the results
of numerical simulation
in Figure 6 for PS2 and in Figure 7 for PS3.
These figures demonstrate good agreement between the
observations and the results of numerical analysis
at all strain rates under consideration.
It is worth noting that the model can correctly
reproduce not only stress overshoot, but also
weak stress undershoot revealed in experiments.
The adjustable parameters $A$, $b$ and $G$ are
presented in Figures 3 to 5 as functions of
shear rate together with their approximations
by Eqs. (43) and (79).
The ``best-fit" values of the parameter $r$
are collected in Table 2.

Our aim now is to discuss the effects of strain
rate and molecular weight of polystyrene solutions
on the adjustable parameters in the constitutive
equations.

\section{Discussion}

According to Figure 5, the elastic modulus $G$
grows with strain-rate intensity $D_{\rm i}$
for all solutions.
It follows from Table 2 that the rate of increase
in $G$ [the exponent $\nu_{G}$ in Eq. (43)] monotonically
decreases with number of entanglements per chain
$M_{\rm w}/M_{\rm e}$.
This observation appears to be natural if we treat
changes in $G$ with strain rate within the Doi--Edwards
reptation theory as a result of rearrangement of tubes
driven by macro-deformation.
The latter implies that changes in the number of
non-relaxed strands with $\dot{k}$
should be extremely pronounced for a weakly entangled
network
(where a strand is practically the same as a chain),
and should be relatively weak for a strongly entangled
system (where the number of entanglements per chain
is rather large, and only strands located at the ends
of a chain can easily escape from the tubes).

The other parameter that characterizes the effect of
strain rate on the elastic modulus, the coefficient
$K_{G}$ in Eq. (43), appears to correlate
with mass-average molecular weight $M_{\rm w}$:
the higher the molecular weight is, the smaller is
the pre-factor $K_{G}$.
This implies that both factors, the mass-average
molecular weight and the average number of
entanglements per chain strongly affect the
dependence of the shear modulus on strain rate.

Table 2 shows no direct correlations between the
plateau modulus $G_{0}$ and the mass-average molecular
weight $M_{\rm w}$.
It reveals, however, a noticeable dependence of
$G_{0}$ on the ratio $M_{\rm w}/M_{\rm e}$:
the larger the number of strands per chain is,
the higher is the modulus $G_{0}$.
This conclusion is in excellent agreement with
the classical theory of rubber elasticity, which
predicts that the elastic modulus is determined
by the number of strands per unit volume,
and it is independent of the average length of a strand.

According to Figure 3, the parameter $A$ grows
with strain-rate intensity, and the rate of increase
in $A$ with shear rate is practically independent
of the network structure (the exponent $\nu_{A}\approx 0.4$
weakly depends on molecular weight and concentration
of entanglements).
Table 2 demonstrates, however, that
the absolute values of $A$ are noticeably
affected by the ratio $M_{\rm w}/M_{\rm e}$:
the coefficient $K_{A}$ substantially decreases
(by an order of magnitude) with average number
of entanglements per chain.
This finding seems natural if we recall that $A$ (which
is proportional to $\alpha$) characterizes the influence
of the network elasticity on the non-affine flow
of junctions that is described by the tensor ${\bf M}$.
When the number of junctions per unit volume is relatively
large, stresses in strands are not sufficient to
substantially affect the sliding process (which implies
that ${\bf D}_{\rm s}\approx \beta {\bf D}$).
On the contrary, for a network with a relatively small
concentration of junctions, stresses in strands
strongly influence flow of junctions (which means
that the tensor ${\bf M}$ does not vanish, and, as a
consequence, the coefficient $\alpha$ in Eq. (29) is
relatively large).

According to Figure 4, the parameter $b$,
that characterizes the rate of sliding of
junctions driven by macro-deformation,
grows with shear rate at all strain rates
under consideration.
For all solutions, the parameter $b$ remains below
unity.
The rate of increase in $b$ with $\dot{k}$
reaches maximum for PS2 (the lowest $M_{\rm w}$),
whereas these rates are rather modest for PS1
and PS3 (solutions with high molecular weights).
Table 2 shows that the exponent $\nu_{b}$ in Eq. (44)
increases with mass-average molecular weight,
and it reveals no correlations with the average
number of entanglements per chain.

According to Table 2, the coefficient $r$
noticeably decreases with the ratio $M_{\rm w}/M_{\rm e}$.
This implies that the effect of strain energy on
the rate of internal dissipation [characterized by the
coefficient $\alpha$ in Eq. (29)] is substantial for
solutions of weakly entangled polymers and may be
disregarded for strongly entangled networks.

\section{Validation of the model}

To verify the constitutive equations, we apply two
approaches.
According to the first,
we analyze the same deformation mode (shear flow),
but for a noticeably denser system
and compare the conclusions drawn in the previous
section with results of numerical simulation.
Following the other approach, we focus on the response
of a polymer melt at another deformation mode
(extensional flow) and compare predictions of the
governing equations (with the adjustable parameters
found by fitting experimental data under simple shear)
with observations.

{\bf 1.}
We begin with the first approach and formulate
explicitly some assertions that can be drawn
for a dense polymer system from the analysis
of results reported in Figures 3 to 5 and Table 2.
As the network under consideration we chose LDPE melt
(IUPAC A) whose physical properties are summarized
in Table 1 with reference to \cite{MPM81,IWH00}
(the mass-average molecular weight $M_{\rm w}$
and the number-average molecular weight $M_{\rm n}$
of the melt are given in \cite{MPM81};
to assess the average number of junctions along a
chain, we use the graph depicted in Figure 3
of \cite{IWH00} and presume that the number of
junctions exceeds the number of branching points
for the highly branched LDPE melt).
Our choice may be explained by two reasons.
First, the observations in start-up shear tests
on LDPE melt at the temperature $\Theta_{0}=150$~$^{\circ}$C
reported in \cite{Mei72} are accompanying by those
in transient extensional tests on the same polymer
at the same temperature \cite{ML79}, which allows
results of numerical simulation to be compared at
two different deformation modes.
Secondly, these observations have recently been used
to validate the pom--pom model \cite{BMH00}.
Thus, the quality of fitting by the constitutive
equations (42) and the pom--pom model may be collated
on the same set of data.

The following hypotheses are introduced regarding
the adjustable parameters of the melt:
\begin{itemize}
\item[{(i)}]
The modulus $G$ should substantially exceed that for
the solutions at all strain rates (this fact is not
trivial because $G$ is determined by fitting the
entire dependencies of the shear stress on time).
The coefficient $G_{0}$ should be relatively
large, whereas the exponent $\nu_{G}$ in Eq. (43)
should be rather small (these quantities are
determined by the ratio $M_{\rm w}/M_{\rm e}$
that is high for the melt).
The pre-factor $K_{G}$ in Eq. (43) should be
relatively large (this quantity is inversely
proportional to the mass-average molecular weight,
while $M_{\rm w}$ of the melt is small compared with
the molecular weights of PS solutions).

\item[{(ii)}]
The coefficient $r$ in Eq. (47) that describes the effect
of strain energy on the rate of sliding of junctions should
be relatively small
(because the ratio $M_{\rm w}/M_{\rm e}$ of the melt
exceeds that of PS solutions).

\item[{(iii)}]
The exponent $\nu_{A}$ of the melt should be close
to those of PS solutions (this parameter is independent
of the molecular structure).
The coefficient $K_{A}$ in Eq. (46) that accounts for
the influence of strain rate on the rate of sliding
of junctions should be quite small (this quantity
is inversely proportional to the average number of
strands in a chain).

\item[{(iv)}]
The exponent $\nu_{b}$ in Eq. (44) that describes
the effect of strain rate on the parameter $b$
(the latter characterizes
the proportionality between the velocity gradients
for macro-deformation and sliding of junctions)
should be relatively small (because the ratio
$M_{\rm w}/M_{\rm e}$ of the melt exceeds that
of PS solutions).
\end{itemize}

To examine these hypotheses, we begin with the
approximation of the experimental data for the
shear stress $\Sigma_{12}$ plotted as a function
of time in Figure 8.
To reduce the number of material constants,
we set $r=0$ in accord with assumption (ii).
To find the coefficients $A$, $b$ and $G$,
each curve $\Sigma_{12}(t)$ is fitted separately
by using the same approach that was utilized in
Section 7.
We chose some values of $A$, $b$ and $G$ from
the intervals $[0,A_{\max}]$, $[0,b_{\max}]$
and $[0,G_{\max}]$, where the ``best-fit" parameters
are assumed to be located,
integrate Eqs. (66) numerically,
and calculate the shear stress $\Sigma_{12}$ by
means of Eq. (67).
The ``best-fit" coefficients $A$, $b$ and $G$
are found from the condition of minimum of the
cost function $R$ in Eq. (78), where the last term
is disregarded.

Figure 8 demonstrates excellent agreement
between the observations and the results of
numerical simulation.
It is worth noting that the quality of fitting
the experimental data is noticeably higher than
that for the pom--pom model with a substantially
larger number of experimental constants,
compare Figure 8 with Figure 9 in \cite{BMH00}.
By no means, this conclusion implies that Eqs. (42)
are superior with regard to the constitutive
equations in the pom--pom model, as the latter
are grounded on the solid basis of the molecular
theory of polymer networks.
We suppose that the good quality of matching
observations with the smaller number of adjustable
parameters revealed in Figure 8 may be explained
by the fact that a detailed description of
orientation and stretch of chains is excessive
for the analysis of shear flows with relatively
large strain rates.

The modulus $G$ determined in the fitting procedure
is plotted versus strain rate $\dot{k}$ in Figure 5.
The dependence $G(\dot{k})$ is matched by Eq. (43)
with the adjustable parameters $G_{0}$, $K_{G}$
and $\nu_{G}$ reported in Table 2.
Comparison of the results listed in Table 2
shows that the modulus $G$ of the melt
substantially exceeds that for the solutions
(at least, by three orders of magnitude),
the exponent $\nu_{G}$ for the melt is small compared
to that for PS solutions,
whereas the pre-factor $K_{G}$ exceeds that for
the solutions.
All these conclusions are in perfect agreement with
assumption (i).

It is obvious that assumption (ii) is fulfilled, as
the value $r=0$ was set in the fitting procedure.

The coefficient $A$ is plotted versus $\dot{k}$ in
Figure 3 together with its approximation by Eq. (79)
with the coefficients $K_{A}$ and $\nu_{A}$
listed in Table 2.
Comparison of these parameters with appropriate
parameters for PS solutions demonstrates that
assumption (iii) is satisfied.

The parameter $b$ is plotted versus strain-rate
intensity $D_{\rm i}$ in Figure 4.
This figure demonstrates that $b$ is practically
constant, which means that the exponent $\nu_{b}$
in Eq. (44) vanishes, in agreement with assumption
(iv).

This analysis reveals that the approximation of
experimental data in shear tests for polymer solutions
by the constitutive model implies the assumptions
regarding the behavior of material parameters of
a melt that appear to be physically plausible.

{\bf 2.}
To further examine the constitutive equations,
we intend to deduce a relationship
that does not contain adjustable parameters
and that connects some quantities which may be
measured directly in conventional tests.
For this purpose, we confine ourselves to relatively
dense polymer systems (melts), and, based on
assertion (ii), set $r=0$ in the governing equations.
According to Eqs. (65) and (77), in this case the first
normal stress function reads
\begin{equation}
\Psi_{1}=\frac{2G(1-b)^{3}}{A^{2}}p_{20}^{4}.
\end{equation}
Equations (76) and (80) result in
\[
\frac{\Psi_{1}}{\eta_{\rm s}}=\frac{2}{A}(1-b)p_{20}.
\]
On the other hand, Eqs. (65) and (74) (where we set $r=0$)
imply that
\[
\frac{1-b}{A}=\frac{1}{D_{\rm i}}
\sqrt{\frac{1-p_{20}^{3}}{p_{20}^{5}}}.
\]
Combining these equalities, we find that
\[
\frac{1-p_{20}^{3}}{p_{20}^{3}}=\Bigl (
\frac{\Psi_{1}D_{\rm i}}{2\eta_{\rm s}}\Bigr )^{2},
\]
which means that
\begin{equation}
p_{20}=\Bigl [ 1+\Bigl (
\frac{\Psi_{1}D_{\rm i}}{2\eta_{\rm s}}\Bigr )^{2}
\Bigr ]^{-\frac{1}{3}}.
\end{equation}
It follows from Eqs. (59) and (76) that
\[
\eta_{\rm e}=3\eta_{\rm s}p_{20}^{-3}.
\]
Substituting expression (81) into this equality, we
arrive at the formula
\begin{equation}
\eta_{\rm e}=3\eta_{\rm s} \Bigl [ 1+
(\frac{\Psi_{1}D_{\rm i}}{2\eta_{\rm s}})^{2} \Bigr ],
\end{equation}
which expresses the steady elongational viscosity
$\eta_{\rm e}$ in terms of the steady shear viscosity
$\eta_{\rm s}$ and first normal stress function
$\Psi_{1}$ for an arbitrary strain-rate intensity $D_{\rm i}$.

To check the validity of Eq. (82), we use observations
on polypropylene melt ($M_{\rm w}=3.5\cdot 10^{5}$ g/mol)
at $\Theta_{0}=190$~$^{\circ}$C
reported in Figure 10 of \cite{BMT96}.
It should be noted that steady extensional and shear
tests are conventionally performed at different
strain rates, which implies that some results provided
in \cite{BMT96} are extrapolations of experimental data.
The steady elongational viscosity $\eta_{\rm e}$
is plotted versus strain-rate intensity $D_{\rm i}$
in Figure 9 (unfilled circles) together with its
prediction by Eq. (82) (filled circles).
For comparison, we also present the estimate of the
elongational viscosity for a Newtonian fluid
$\eta_{\rm e}=3\eta_{\rm s}$ (asterisks).
Figure 9 shows that Eq. (82) provides a reasonable
correction of the Newtonian formula for the dependence
of the elongational viscosity on strain rate.

To further examine Eq. (82), we consider experimental
data for dilute solutions of polyacrylamide in
water/glycerol solvent \cite{PVM03}.
The observations depicted in Figures 4 and 6 of \cite{PVM03}
show that the steady shear viscosity $\eta_{\rm s}$
and the first normal stress coefficient $\Psi_{1}$
remain practically constant at shear rates up to
200--500 s$^{-1}$ (the low-shear plateau).
Formula (82) predicts that in this region of strain
rates, the steady elongational viscosity $\eta_{\rm e}$
should increase with strain-rate intensity (strain-thickening).
This conclusion is in agreement with the data depicted
in Figure 10 of \cite{PVM03}, which show a pronounced
increase (by an order of magnitude) in the
elongational viscosity with $\dot{\epsilon}$ in the
interval of strain rates from 1 to 200 s$^{-1}$.
We do not provide qualitative comparison of the
model predictions with these observations (as it was
mentioned before, the assumption $r=0$ is not necessary
fulfilled for polymer solutions,
whereas an analog of Eq. (82) for $r>0$ becomes too
cumbersome).
However, it is worth noting that constitutive equations
(42) can describe (at least, qualitatively)
strain-thickening in polymer fluids.
This conclusion is rather surprising, because the
strain-thickening phenomenon is conventionally associated
with coil-stretch transition in polymer solutions, while
the latter effect is not accounted for by the model
explicitly.

{\bf 3.}
Our aim now is to show that Eqs. (42) can correctly
describe the experimental data on elongational flows
with constant strain rates (in particular,
strain-hardening observed in uniaxial elongational
tests) when their parameters are found by matching
observations in shear tests.
For this purpose, we study the evolution of the
transient viscosity $\eta_{\rm e}^{+}$
with time for the same LDPE melt that was used in
the analysis of stress overshoot in shear tests.
The experimental data in start-up elongational
tests at $\Theta_{0}=150$~$^{\circ}$C reported
in \cite{ML79} are depicted in Figure 10.

It is worth noting a pronounced difference between the
observations in extensional and shear tests presented
in Figures 8 and 10.
While the stress overshoot and subsequent strain
softening in shear tests (Figure 8) occur at times
of the order of 1 to 10 s (at the chosen shear rates),
a pronounced strain hardening is observed in extensional
tests at times of the order of 10$^{-2}$ to 10$^{-1}$ s.
In this time interval, our assumption regarding rapid
tuning of the number of strands to its steady value
corresponding to a given strain rate becomes
inadequate, and the evolution of the function $N(t)$
should be taken into account.

Within the reptation concept, the quantity $N(t)$
equals the concentration of tubes survived until
instant $t\geq 0$.
It was demonstrated in \cite{DE86} that changes in
the number of survived tubes in a matrix of fixed
obstacles are correctly described by a single-exponential
function.
This implies that, as a first approximation,
the first-order kinetic equation can be introduced
for the function $N(t)$,
\[
\frac{dN}{dt}(t)=\frac{1}{\tau_{\rm r}(D_{\rm i})}
\Bigl [ N^{0}(D_{\rm i}) -N(t)\Bigr ],
\]
where $N^{0}(D_{\rm i})$ is the steady concentration
of strands at deformation with the strain-rate intensity
$D_{\rm i}$,
and $\tau_{\rm r}(D_{\rm i})$ is the characteristic time
for changes in the number of strands at transition from
the rest to the steady flow.
As the elastic modulus $G$ is proportional to $N$,
we find from this equation that the function $G(t)$
obeys the relation
\begin{equation}
\frac{d G}{dt}(t)=\frac{1}{\tau_{\rm r}(D_{\rm i})}
\Bigl [G^{0}(D_{\rm i}) -G(t)\Bigr ],
\end{equation}
where $G^{0}(D_{\rm i})$ is given by Eq. (43).

After this correction of the governing equations
driven by the necessity to account for the evolution
of the elastic modulus, the transient viscosity
$\eta_{\rm e}^{+}$ in a start-up elongational test
with a constant strain rate $\dot{\epsilon}$
is determined by Eq. (58), where the stress difference
$\Delta\Sigma$ obeys Eq. (54), the function $\kappa(t)$
is governed by Eq. (53), and the coefficient $G$
is described by Eq. (83).

The following procedure is applied to fit the
observations at various strain rates $\dot{\epsilon}$
depicted in Figure 10.
First, we set $r=0$ and calculate $G^{0}(\dot{\epsilon})$
and $A(\dot{\epsilon})$ by using Eqs. (43) and (79)
[where $D_{i}$ is given by Eq. (52)] with the parameters
$G_{0}$, $K_{G}$, $\nu_{G}$, $K_{A}$ and $\nu_{A}$
found in the approximation of observations in shear tests.
As the initial condition $G_{\rm in}$ for Eq. (83)
is unknown
[it corresponds to the elastic modulus after sudden
application of the external load and does not necessary
coincide with $G_{0}$ in Eq. (43)],
we begin with matching observations for the elongational
viscosity in a test with the smallest strain rate
$\dot{\epsilon}=0.3$ s$^{-1}$
(for this test, the largest number of experimental data
are presented in Figure 10).
The curve $\eta_{\rm e}^{+}(t)$ is determined by 3
constants: $G_{\rm in}$, $b$ and $\tau_{\rm r}$
(formally, the coefficient $b$ found in the shear tests
can be utilized, but we chose to treat $b$
as an adjustable parameter in order to compare appropriate
results in extensional and shear tests:
the closeness of values of $b$ found in two different
tests will confirm that the fitting procedure is stable).
The quantities $G_{\rm in}$, $b$ and $\tau_{\rm r}$
are determined by the same algorithm that was employed
in Section 7.
We fix some intervals, where the ``best-fit" parameters
are located, divide these intervals into subintervals,
for each triad chosen from the subintervals integrate
Eqs. (53) and (83) numerically (by the Runge--Kutta
method with the time-step $\Delta t=1.0\cdot 10^{-5}$ s),
and calculate $\eta_{\rm e}^{+}$ from Eqs. (54) and (58).
The best-fit values of $G_{\rm in}$, $b$ and $\tau_{\rm r}$
are chosen from the condition of minimum of the cost
function
\[
R=\sum_{t_{m}} \Bigl [ \eta_{\rm e}^{\rm exp}(t_{m})
-\eta_{\rm e}^{\rm num}(t_{m}) \Bigr ]^{2},
\]
where the sum is calculated over all experimental points
$t_{m}$ depicted in Figure 15,
$\eta_{\rm e}^{\rm exp}$ is the elongational viscosity
measured in the test,
and $\eta_{\rm e}^{\rm num}$ is given by Eq. (58).

The initial modulus $G_{\rm in}\approx 2.0\cdot 10^{6}$ Pa
ensures the best approximation of the experimental data
in the extensional test with the strain rate
$\dot{\epsilon}=0.3$ s$^{-1}$.
We now fix this value and fit other curves depicted
in Figure 10 by using the same algorithm with
only two adjustable parameters, $b$ and $\tau_{\rm r}$.
Each set of observations is matched separately.

Figure 10 demonstrates quite reasonable agreement between
the experimental data and the results of numerical
simulation at all strain rates under consideration.
The adjustable parameter $b$ is plotted versus strain-rate
intensity $D_{\rm i}$ in Figure 11.
Comparison of the dependencies $b(D_{\rm i})$ found
by matching observations in shear (unfilled circles)
and extensional (filled circles) tests
reveals that the experimental values of $b$ practically
coincide, which implies that the fitting algorithm
is stable.

The parameter $\tau_{\rm r}$ is plotted versus
strain-rate intensity $D_{\rm i}$ in Figure 12
(unfilled circles).
According to this figure, the parameter $\tau_{\rm r}$
strongly decreases with strain-rate intensity $D_{\rm i}$.
The experimental data are approximated by the
phenomenological relation
\begin{equation}
\log \tau_{\rm r}=\tau_{\rm r0}-\tau_{\rm r1}\log D_{\rm i},
\end{equation}
where the coefficients $\tau_{\rm r0}$ and $\tau_{\rm r1}$
are calculated by the least-squares method.
Figure 12 shows that Eq. (84) ensures fair approximation
of the observations.
For comparison, we also present in Figure 12 the instants
$\tau_{\rm p}$, when the stress $\Sigma_{12}$ in shear
tests reaches its maximal values (filled circles).
The observations are fitted by the formula similar to
Eq. (84),
\begin{equation}
\log \tau_{\rm p}=\tau_{\rm p0}-\tau_{\rm p1}\log D_{\rm i}
\end{equation}
with the coefficients $\tau_{\rm p0}$ and $\tau_{\rm p1}$
found by the least-squares technique.
Figure 12 demonstrates that at all strain rates under
consideration,
the peak instants $\tau_{\rm p}$ substantially
(at least, by an order of magnitude) exceed the relaxation
times $\tau_{\rm p}$.
This confirms our assumption that the evolution of the
elastic modulus $G$ with time can be disregarded in
the analysis of stress overshoot under shear.

{\bf 4.}
As the objective of this work is to study the rate-dependent
behavior of polymer fluids at large deformations,
we do not concentrate on their response in oscillatory
tests with small strains and relaxation tests.
It is instructive, however, to make some comments
regarding the model predictions in conventional
viscoelastic tests.

We begin with the analysis of governing equations at
small strains.
For this purpose, we set
\begin{equation}
{\bf B}_{\rm e}={\bf I}+2\hat{\epsilon}_{\rm e},
\end{equation}
where $\hat{\epsilon}$ is the Finger strain tensor
for elastic deformation,
substitute expression (86) into Eqs. (42),
and neglect terms of the second order of smallness with
respect to $\hat{\epsilon}_{\rm e}$ and $\hat{\epsilon}$,
where $\hat{\epsilon}$ is the strain tensor for
macro-deformation.
Using Eqs. (43) to (47), we arrive at the expression for
the stress tensor,
\begin{equation}
{\bf \Sigma}(t)=-P(t){\bf I}+2G_{0}\hat{\epsilon}_{\rm e}(t),
\end{equation}
where the same notation $P$ is employed for the unknown
pressure.
The evolution of the elastic strain tensor is described
by the differential equation
\begin{equation}
\frac{d\hat{\epsilon}_{\rm e}}{dt}(t)+
\frac{1}{\tau_{\rm s}}\hat{\epsilon}_{\rm e}(t)
=\frac{d\hat{\epsilon}}{dt}(t),
\qquad
\tau_{\rm s}=\frac{1}{A_{0}}.
\end{equation}
Formulas (87) and (88) reveal that Eqs. (42) are
transformed into the Maxwell model at small strains,
which implies that these relations can
adequately describe observations in conventional
oscillatory tests,
provided that the one-mode model (42) is replaced
by its multi-mode analog.
As this procedure is straightforward, we do not
dwell on it in the present study.

Two comments are noteworthy.
First, the fact that the elastic modulus at
small strains coincides with $G_{0}$ provides
an additional opportunity to verify
the model predictions, because $G_{0}$ can be
measured independently in shear relaxation
tests, on the one hand, and it can be calculated
from Eq. (43) by fitting the experimental data
on $G(\dot{k})$ in start-up shear tests, on the
other.
Although we do not expect this procedure to result
in the same values of $G_{0}$ for LDPE melt
[we have only 4 data points for $G(\dot{k})$ that
are approximated by the three-parameter formula (43),
which means that the accuracy of determining $G_{0}$
is rather poor],
we would mention that the value
$G_{0}^{\rm start-up}=6.5\cdot 10^{4}$ Pa
reported in Table 2 overestimates by less than twice
the value $G_{0}^{\rm rel}=3.4\cdot 10^{4}$ Pa
evaluated from the data in relaxation tests
with small strains \cite{Mei72}.

Secondly, the relaxation time $\tau_{\rm s}$ in Eq. (88)
is inversely proportional to the parameter $A_{0}$.
To derive constitutive equations that are equivalent
to the Maxwell model at small strains was the main
reason to preserve the first term in phenomenological
equation (46) despite the fact that all observations
in the interval of strain rates under consideration
can be approximated quite well by Eq. (79) without $A_{0}$.
Following Eq. (88), we can introduce an effective
relaxation time $\tau_{\rm s}$ for an arbitrary
strain-rate intensity $G_{\rm i}$ as the reciprocate
of the function $A(D_{\rm i})$.
The experimental data for $\tau_{\rm s}(D_{\rm i})$
are depicted in Figure 12 together with their
approximation by the equation analogous to
Eqs. (84) and (85),
\begin{equation}
\log \tau_{\rm s}=\tau_{\rm s0}-\tau_{\rm s1}\log D_{\rm i}.
\end{equation}
The coefficients $\tau_{\rm s0}$ and $\tau_{\rm s1}$
in Eq. (89) are found by the least-squares technique.
Figure 12 shows that the slopes of the curves
$\tau_{\rm r}(D_{\rm i})$ and $\tau_{\rm s}(D_{\rm i})$
are quite similar, and for all strain rates,
the effective relaxation time $\tau_{\rm s}$ exceeds
the relaxation time for the elastic modulus $\tau_{\rm r}$
by two orders of magnitude.
This provides another confirmation of our hypothesis
that the evolution of the elastic modulus $G$ to its
ultimate value $G^{0}(D_{\rm i})$ may be neglected
in the study of start-up tests.
Figure 12 also shows that given a strain rate,
the effective relaxation time $\tau_{\rm s}$
is substantially higher than the time for stress
overshoot $\tau_{\rm p}$, which means that $\tau_{\rm p}$
cannot be employed to assess the function $A(D_{\rm i})$,
and the entire curves $\Sigma_{12}(t)$ should be
approximated to determined the model parameters.

Finally, we consider relaxation tests with finite
deformations.
In these experiments, constitutive equations (42) with
necessary correction (83) demonstrate a
time-dependent behavior with two characteristic
time-scales:
(i) rapid relaxation driven by changes in the
modulus $G$ and (ii) slow decrease in stress
associated with the Maxwell-type evolution of
the elastic Cauchy--Green tensor ${\bf B}_{\rm e}$
in Eqs. (42).
This implies that the stress in a step-strain
test cannot be factorable into a product of
a strain-dependent and time-dependent functions.
As the same features are also revealed by the
pom--pom model, we do not discuss them in detail,
referring to \cite{ML98}, where recent
observations data are analyzed that confirm
the presence of two time-scales in the viscoelastic
response of polymer fluids.

\section{Concluding remarks}

A constitutive model has been developed for a polymer
fluid.
A polymer is thought of as an incompressible network
of strands bridged by junctions.
The network is treated as permanent (strands
cannot separate from their nodes).
The junctions between strands can slide with respect
to their reference positions under deformation.
No restrictions are imposed on the rate-of-strain tensor
for sliding of junctions, whereas the vorticity tensor
for sliding is proportional to that for macro-deformation.
Stress--strain relations are derived by using the laws
of thermodynamics.
For a Gaussian network of strands, the constitutive
equations involve three adjustable functions,
for which phenomenological relations are introduced.

In a broad sense, the constitutive equations may be
treated as a combination of the Johnson--Segalman
and Leonov models.
The important advantages of these relations
are that (i) they are based on the conventional
multiplicative decomposition of the deformation
gradient (no slip tensors are used),
and (ii) they presume (in agreement with the classical
theories in polymer physics) the strain energy of
strands to be Gaussian (no complicated formulas are
introduced for the elastic energy).

The governing equations are simplified for uniaxial
extension and simple shear with finite strains.
Explicit formulas are developed for the steady
elongational and shear viscosities, as well as for
the normal stress functions.

To verify the model, three sets of experimental
data are approximated for monodisperse polystyrene
solutions with various molecular weights and
molecular weights between entanglements.
Good agreement is demonstrated between the
observations for stress overshoot in start-up
shear tests and the results of numerical simulation
at various strain rates.
It is revealed that the phenomenological relations
correctly characterize the effect of strain rate on
the material parameters, and their coefficients
change consistently with mass-average molecular
weight and concentration of entanglements.

To validate the governing equations, the evolution
of stresses is studied in start-up extensional
and shear flows on low-density polyethylene melt.
It is demonstrated that constitutive equations (42)
and (83) adequately describe changes in the
elongation viscosity with time when their parameters
are found by matching observations in shear tests.

Our approximation of observations in extensional and shear
flows of LDPE melt demonstrates higher quality of
fitting compared to the pom--pom model.
An advantage of constitutive equations (42) is that they
can also be used to match the experimental data
on dilute polymer solutions, a class of polymer
fluids to which the pom--pom model is inapplicable.

New formula (82) has been derived that expresses
the steady elongational viscosity in terms of the
steady shear viscosity and the first normal stress
function.
The validity of this equation is confirmed by
observations on polypropylene melt.

The following characteristic features of the constitutive
equations have been revealed:
\begin{enumerate}
\item
The model can quantitatively describe overshoots
for the shear stress and the first normal stress
difference in polymer melts and solutions, as well as
stress undershoot in polymer solutions.

\item
It can quantitatively describe strain hardening in
start-up extensional tests.

\item
The governing equations can qualitatively predict
strain-thickening in steady extensional tests.

\item
The model can qualitatively describe two-scale
stress relaxation in polymer fluids in step-strain
tests.

\item
At small strains, the constitutive equations are
transformed into the conventional Maxwell model.
\end{enumerate}

\newpage
\subsection*{List of tables}

\begin{description}

\item{
{\bf Table 1:}
 Concentrations $c$,
mass-average molecular weights $M_{\rm w}$,
polydispersity indices $D$,
and average numbers of entanglements per chain
$M_{\rm w}/M_{\rm e}$ for PS solutions and LDPE melt.}
\vspace*{-1 mm}

\item{
{\bf Table 2:}
Adjustable parameters for LDPE melt and PS solutions.}
\end{description}
\vspace*{20 mm}

\begin{center}
{\bf Table 1:} Concentrations $c$,
mass-average molecular weights $M_{\rm w}$,
polydispersity indices $D$,
and average numbers of entanglements per chain
$M_{\rm w}/M_{\rm e}$ for PS solutions and LDPE melt
\vspace*{6 mm}

\begin{tabular}{@{} l c c c c l @{}}\hline
Abbreviation & $c$            & $M_{\rm w}$ g/mol  & $D$  & $M_{\rm w}/M_{\rm e}$ & Reference \cr \hline
PS1          & 3.0 wt-\%      & $8.42\cdot 10^{6}$ & 1.17 & 10.0                  & \cite{OLM98} \\
PS2          & 0.1 g/cm$^{3}$ & $1.09\cdot 10^{6}$ &      & 6.1                   & \cite{OII00}  \\
PS3          & 0.1 g/cm$^{3}$ & $8.24\cdot 10^{6}$ &      & 1.8                   & \cite{OIU00a} \\
LDPE         &                & $2.47\cdot 10^{5}$ & 15.2 & over 60               & \cite{MPM81,IWH00} \\
\hline
\end{tabular}
\end{center}
\vspace*{30 mm}

\begin{center}
{\bf Table 2:} Adjustable parameters for LDPE melt and PS solutions
\vspace*{6 mm}

\begin{tabular}{@{} c c c c c c c c c @{}}\hline
Abbreviation & $\log K_{A}$ & $\nu_{A}$ & $\log K_{b}$ & $\nu_{b}$ & $G_{0}$ Pa & $K_{G}$ & $\nu_{G}$ & $r$ \cr \hline
LDPE         & $-1.09$ & 0.40 & $-0.06$ & 0.0  & $6.5\cdot 10^{4}$ & 0.172 & 0.11 & 0.0 \\
PS1          & $-0.77$ & 0.40 & $-0.75$ & 0.0  & 79.7               & 0.058 & 0.34 & 0.1 \\
PS2          & $-0.37$ & 0.31 & $-0.78$ & 0.60 & 59.6               & 0.144 & 0.51 & 0.2 \\
PS3          & $-0.15$ & 0.37&  $-0.32$ & 0.15 & 22.0               & 0.082 & 0.91 & 0.4 \\
\hline
\end{tabular}
\end{center}

\newpage
\subsection*{List of figures}

\begin{description}

\item{
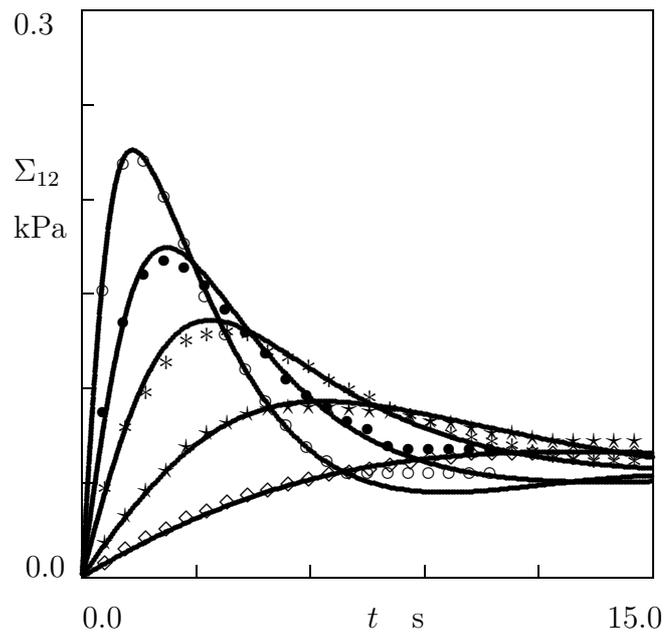
{\bf Figure 1:}
The shear stress $\Sigma_{12}$ versus time $t$.
Symbols: experimental data on PS1 in shear tests
with the strain rates
$\dot{k}=4.0$ (unfilled circles),
2.0 (filled circles),
1.0 (asterisks),
0.5 (stars)
and 0.2 s$^{-1}$ (diamonds) \cite{OLM98}.
Solid lines: results of numerical simulation.}
\vspace*{-1 mm}

\item{
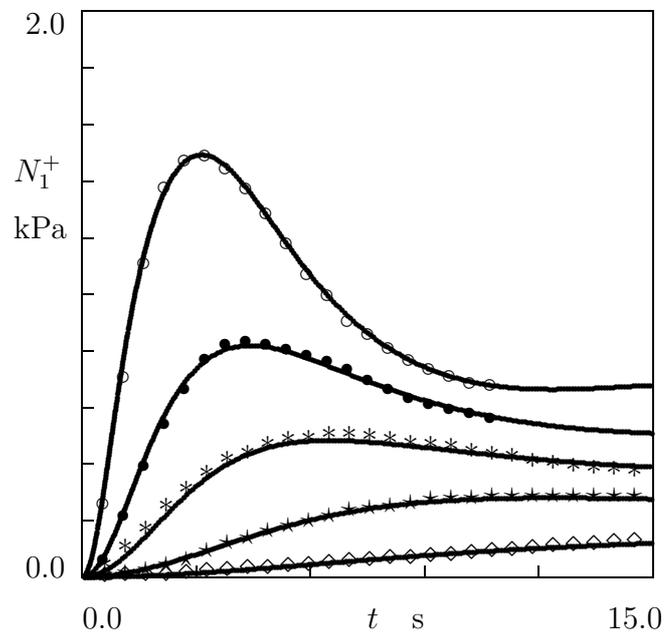
{\bf Figure 2:}
The first normal stress difference $N_{1}^{+}$
versus time $t$.
Symbols: experimental data on PS1 in shear tests
with the strain rates
$\dot{k}=4.0$ (unfilled circles),
2.0 (filled circles),
1.0 (asterisks),
0.5 (stars)
and 0.2 s$^{-1}$ (diamonds) \cite{OLM98}.
Solid lines: results of numerical simulation.}
\vspace*{-1 mm}

\item{
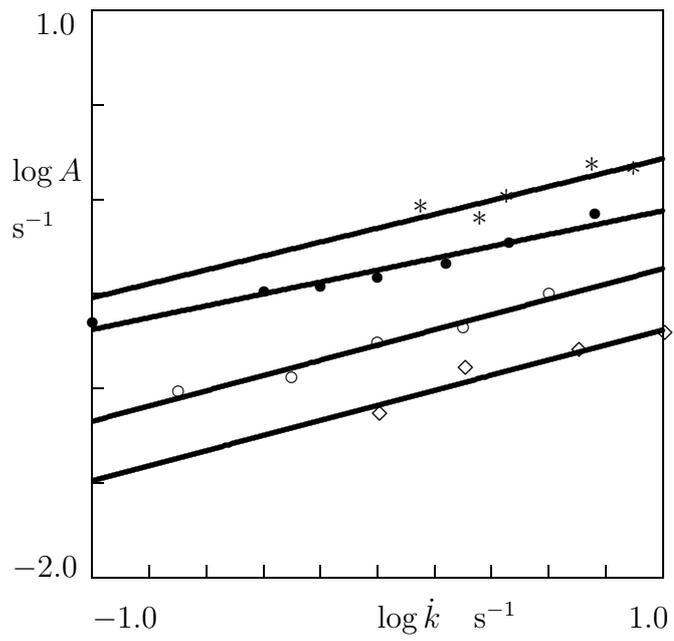
{\bf Figure 3:}
The parameter $A$ versus shear rate $\dot{k}$.
Symbols: treatment of observations on
solutions PS1 (unfilled circles),
PS2 (filled circles),
PS3 (asterisks),
and LDPE melt (diamonds).
Solid lines: approximation of the experimental data
by Eq. (79).}
\vspace*{-1 mm}

\item{
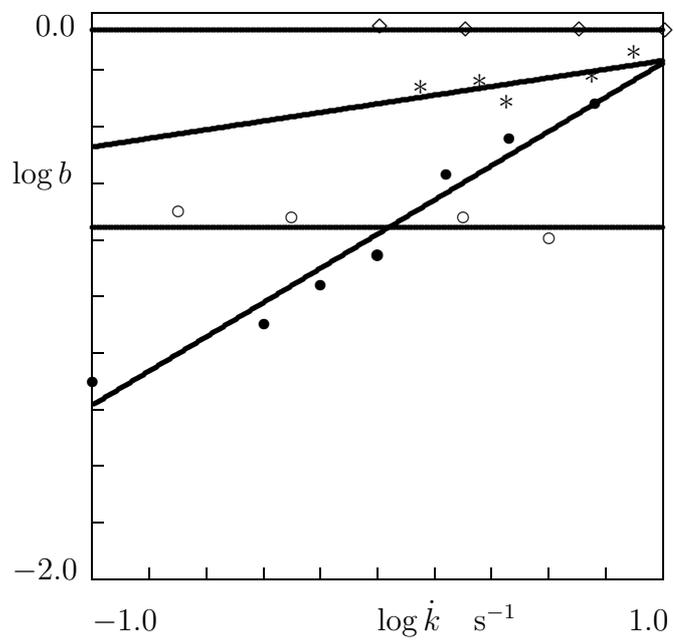
{\bf Figure 4:}
The dimensionless parameter $b$ versus shear rate
$\dot{k}$.
Symbols: treatment of observations on
solutions PS1 (unfilled circles),
PS2 (filled circles),
PS3 (asterisks),
and LDPE melt (diamonds).
Solid lines: approximation of the experimental data
by Eq. (79).}
\vspace*{-1 mm}

\item{
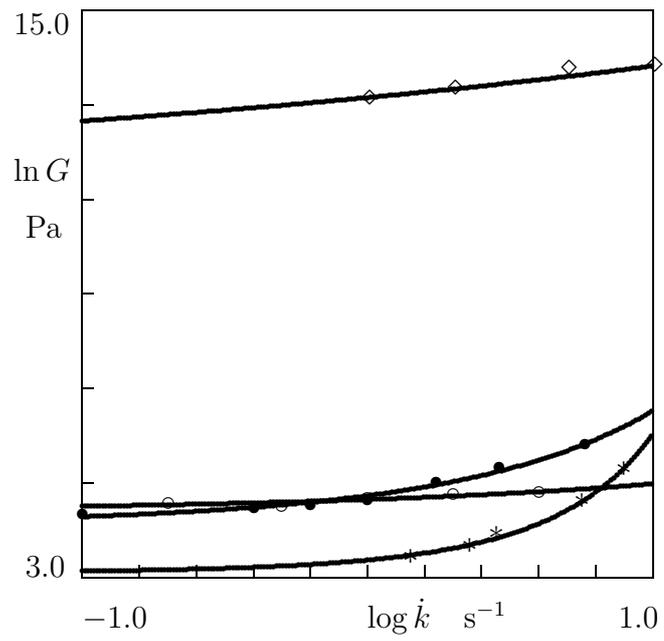
{\bf Figure 5:}
The elastic modulus $G$ versus shear rate $\dot{k}$.
Symbols: treatment of observations on
solutions PS1 (unfilled circles),
PS2 (filled circles),
PS3 (asterisks),
and LDPE melt (diamonds).
Solid lines: approximation of the experimental data
by Eq. (43).}
\vspace*{-1 mm}

\item{
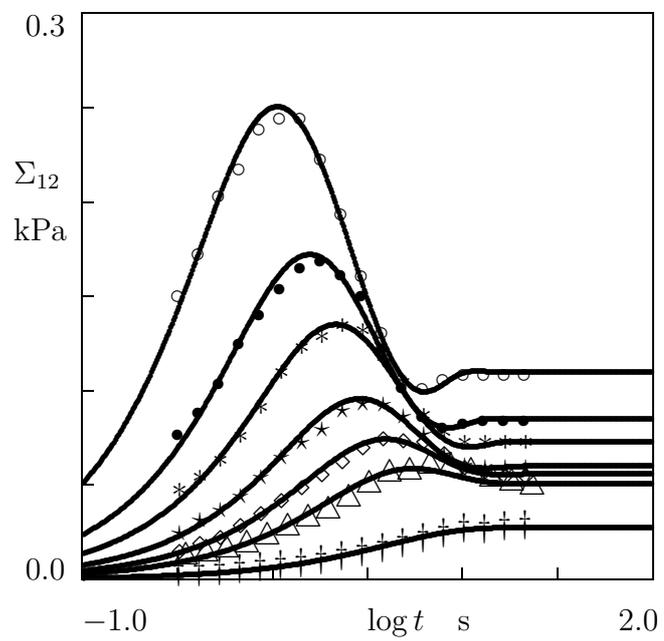
{\bf Figure 6:}
The shear stress $\Sigma_{12}$ versus time $t$.
Circles: experimental data on PS2 in shear tests
with the strain rates
$\dot{k}=5.8$ (unfilled circles),
2.9 (filled circles),
1.74 (asterisks),
1.0 (stars),
0.63 (diamonds),
0.4 (triangles)
and 0.1 s$^{-1}$ (daggers) \cite{OII00}.
Solid lines: results of numerical simulation.}
\vspace*{-1 mm}

\item{
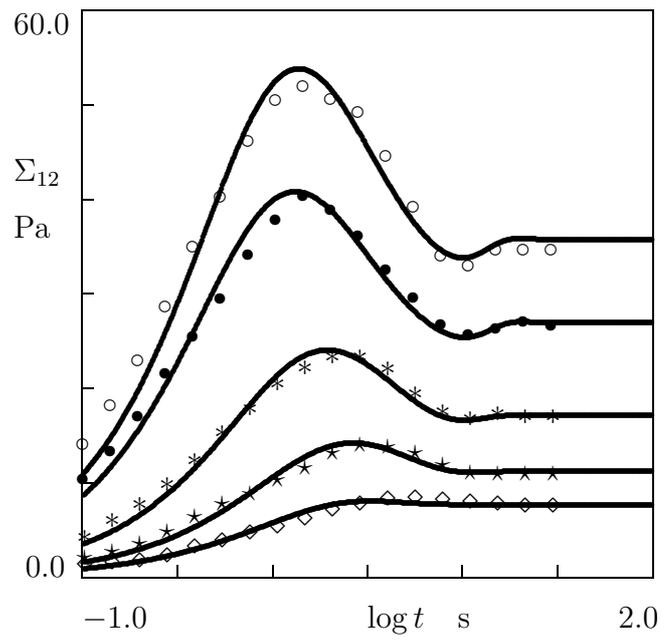
{\bf Figure 7:}
The shear stress $\Sigma_{12}$ versus time $t$.
Symbols: experimental data on PS3 in shear tests
with the strain rates
$\dot{k}=7.77$ (unfilled circles),
5.55 (filled circles),
2.78 (asterisks),
2.24 (stars)
and 1.39 s$^{-1}$ (diamonds) \cite{OIU00a}.
Solid lines: results of numerical simulation.}
\vspace*{-1 mm}

\item{
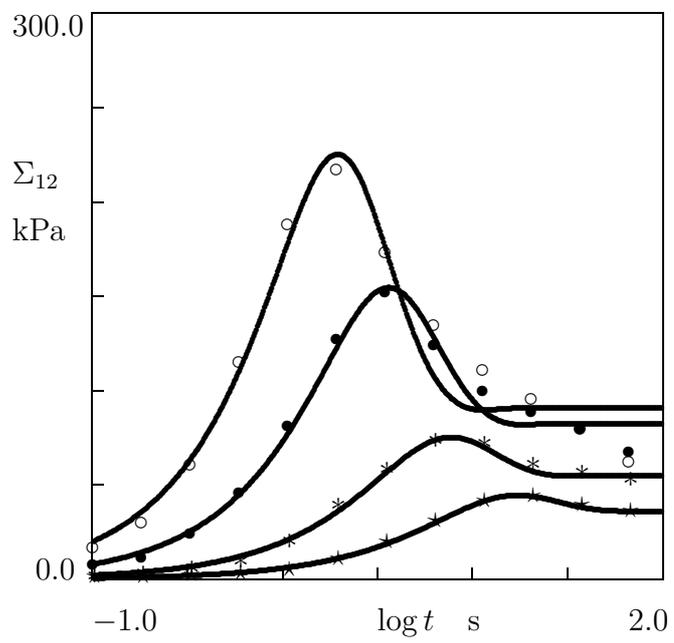
{\bf Figure 8:}
The shear stress $\Sigma_{12}$ versus time $t$.
Symbols: experimental data on LDPE melt in shear
tests with the strain rates
$\dot{k}=10.0$ (unfilled circles),
5.0 (filled circles),
2.0 (asterisks),
and 1.0 s$^{-1}$ (stars) \cite{Mei72}.
Solid lines: results of numerical simulation.}
\vspace*{-1 mm}

\item{
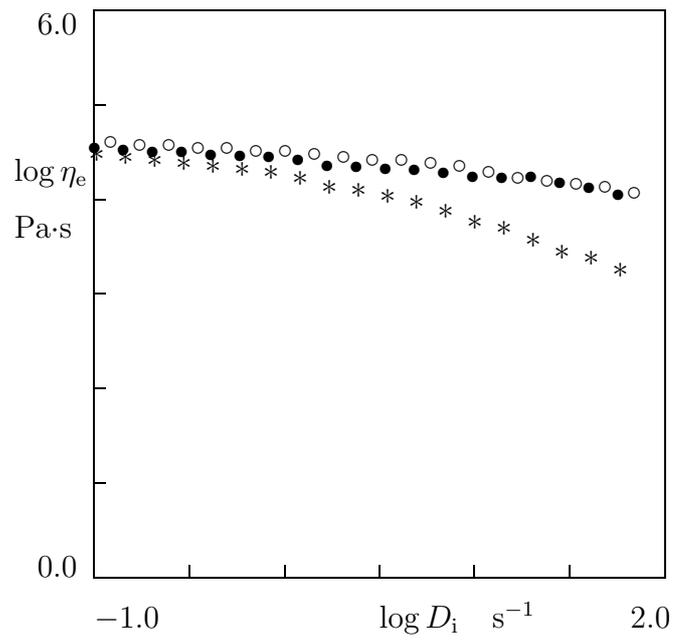
{\bf Figure 9:}
The steady elongational viscosity $\eta_{\rm e}$
versus strain-rate intensity $D_{\rm i}$.
Unfilled circles: experimental data on polypropylene melt
in uniaxial extensional tests  \cite{BMT96}.
Filled circles: predictions of Eq. (82).
Asterisks: the Newtonian estimate $\eta_{\rm e}=3\eta_{\rm s}$.}
\vspace*{-1 mm}

\item{
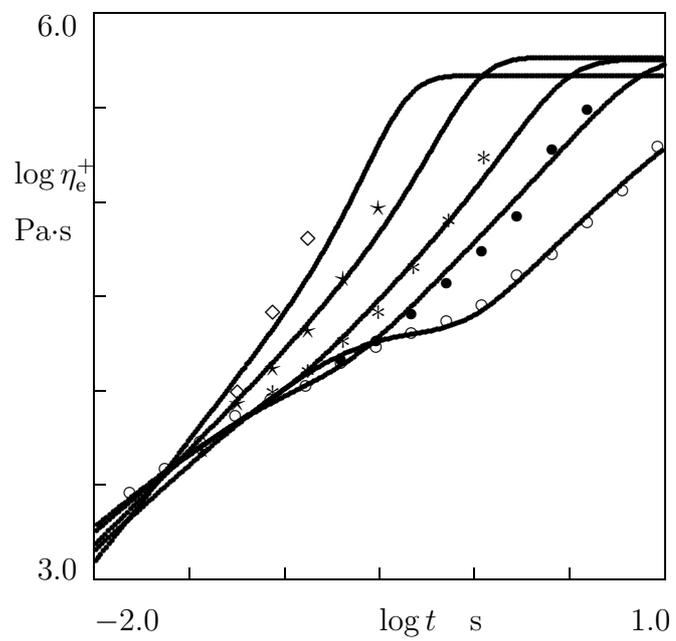
{\bf Figure 10:}
The transient elongational viscosity $\eta_{\rm e}^{+}$
versus time $t$.
Symbols: experimental data on LDPE melt
in uniaxial extensional tests with the strain rates
$\dot{\epsilon}=0.3$ (unfilled circles),
1.0 (filled circles),
3.0 (asterisks),
10.0 (stars)
and 30 s$^{-1}$ (diamonds) \cite{ML79}.
Solid lines: results of numerical simulation.}
\vspace*{-1 mm}

\item{
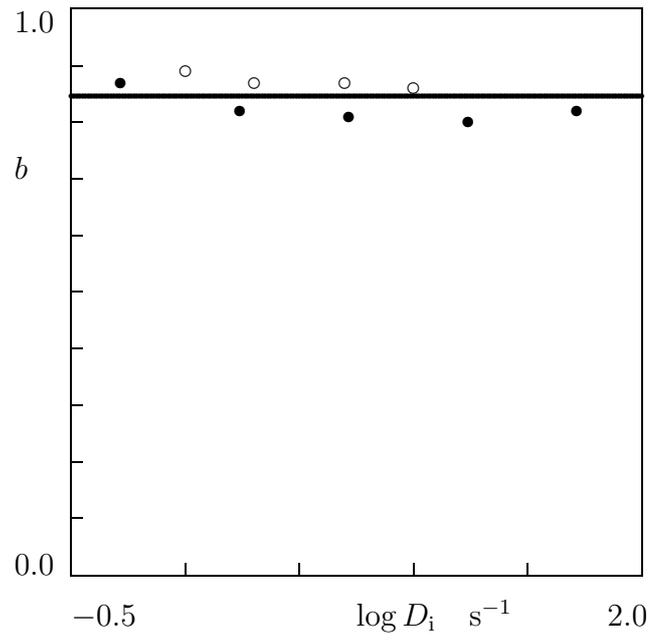
{\bf Figure 11:}
The dimensionless parameter $b$
versus strain-rate intensity $D_{\rm i}$.
Symbols: treatment of observations on LDPE melt
in shear (unfilled circles) and extensional
(filled circles) tests.
Solid line: approximation of the experimental data
by the constant $b=0.85$.}
\vspace*{-1 mm}

\item{
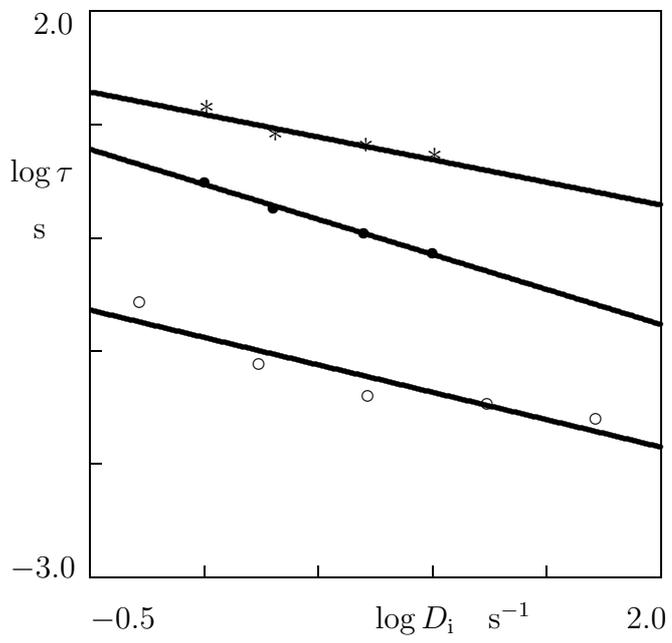
{\bf Figure 12:}
The relaxation time for the elastic modulus
$\tau=\tau_{\rm r}$ (unfilled circles),
the time for stress overshoot $\tau=\tau_{\rm p}$
(filled circles)
and the effective relaxation time for sliding
of junctions $\tau=\tau_{\rm s}$ (asterisks)
versus strain-rate intensity $D_{\rm i}$.
Symbols: treatment of observations on LDPE melt
in shear and extensional tests.
Solid lines: approximation of the experimental data
by Eqs. (84), (85) and (89)
with
$\tau_{\rm r0}=-0.88$, $\tau_{\rm r1}=0.48$,
$\tau_{\rm p0}=0.47$, $\tau_{\rm p1}=0.62$,
and
$\tau_{\rm s0}=0.40$, $\tau_{\rm s1}=1.09$.}
\end{description}

\end{document}